\documentstyle[twocolumn,prl,aps]{revtex}
\input epsf

\begin{document}

\draft
\twocolumn[\hsize\textwidth\columnwidth\hsize\csname@twocolumnfalse\endcsname

\title{Spatial correlations of mobility and immobility in a
glassforming Lennard-Jones liquid}

\author{Claudio Donati$^1$, Sharon C. Glotzer$^1$, Peter H. Poole$^2$, Walter Kob$^3$, and 
Steven J. Plimpton$^4$}

\address{$^1$ Polymers Division and Center for Theoretical and
Computational Materials Science, NIST, Gaithersburg, Maryland, USA 20899}

\address{$^2$Department of Applied Mathematics,
University of Western Ontario, London, Ontario N6A~5B7, Canada}

\address{$^3$ Institut f\"ur Physik, Johannes Gutenberg Universit\"at,
Staudinger Weg 7, D-55099 Mainz, Germany}

\address{$^4$ Parallel Computational Sciences Department,
Sandia National Laboratory, Albuquerque, NM 87185-1111}

\date{\today}
\maketitle

\begin{abstract}
Using extensive molecular dynamics simulations of an equilibrium,
glass-forming Lennard-Jones mixture, we characterize in detail the
local atomic motions. We show that spatial correlations exist among
particles undergoing extremely large (``mobile'') or extremely small
(``immobile'') displacements over a suitably chosen time interval.
The immobile particles form the cores of relatively compact clusters,
while the mobile particles move cooperatively and form
quasi-one-dimensional, string-like clusters.  The strength and length
scale of the correlations between mobile particles are found to grow
strongly with decreasing temperature, and the mean 
cluster size appears to diverge at the mode-coupling critical
temperature.  We show that these correlations in the particle
displacements are related to equilibrium fluctuations in the local
potential energy and local composition.
\end{abstract}

\pacs{PACS numbers: 02.70.Ns, 61.20.Lc, 61.43.Fs}

\vskip2pc]
\narrowtext

\section{Introduction}
\label{sec:intro}

The bulk dynamical properties of many cold, dense liquids differ
dramatically from what might be expected from extrapolation of their
high temperature behavior\cite{science}.  For example, many liquids
cooled below their melting temperature exhibit rapid non-Arrhenius increases
of viscosity and relaxation time with decreasing temperature, and
two-step, stretched exponential decay of the intermediate scattering
function $F({\bf q},t)$. Such behavior is often discussed as a
``signature'' of the approach to the glass transition.  It has long
been a central goal of theories of the glass transition to account for
these bulk phenomena in terms of the microscopic dynamical motions of
the molecules of the liquid.  As a consequence, computer simulations
of supercooled liquids, in which this microscopic information is
immediately available, are increasingly used to complement theoretical
and experimental efforts. In particular, simulations in recent years
have focused on the study of ``dynamical heterogeneity'' to understand
the microscopic origin of slow dynamics and stretched exponential
relaxation in glass-forming liquids
\cite{hiwatari,mountain95,hurley,onuki,doliwa}.

Recently we reported the observation of dynamical heterogeneity
\cite{kdppg} and also cooperative molecular motion \cite{ddkppg} in
extensive molecular dynamics simulations of a supercooled
Lennard-Jones (LJ) mixture. These spatially correlated dynamics were
observed in a regime of temperature $T$, density $\rho$ and pressure
$P$ for $T$ above the dynamical critical temperature $T_c$ obtained
\cite{kobandersen,kob} from fits by the ideal mode coupling theory
(MCT) \cite{mct} to other data on the same system. 
The principle goals of the present paper
are twofold: (1) To test directly for spatial correlations of
particles assigned (according to their displacement over some time) to
subsets of extreme mobility or immobility, and (2) to establish
connections between this ``dynamical heterogeneity'' and local
structure.

This paper is organized as follows. In Section~\ref{sec:background} we
present relevant background information, and in
Section~\ref{sec:model} we describe the model and computer simulation
techniques.  In Sec.~\ref{sec:structure}, we examine the bulk dynamics
and equilibrium structure of the liquid.  In Sec.~\ref{sec:dynamics}
we examine the mean square displacement and analyze the shape of the
time-dependent distribution of particle displacements to define a time
scale which we use to study dynamical heterogeneity throughout the
remainder of the paper.  In Sec.~\ref{sec:mobility} we group particles
into subsets according to the maximum displacement they achieve on the
time scale defined in the previous section, and show that particles of
extremely high or low displacement are spatially correlated.  From
this we are able to identify a length scale that grows with
decreasing $T$.  In Sec.~\ref{sec:energy} we show that fluctuations of
the local mobility are correlated to fluctuations of the potential
energy, or alternatively to fluctuations in the local composition of
the liquid. In Sec.~\ref{sec:lifetime} we examine certain time
dependent quantities associated with the observed dynamical
heterogeneity, and finally in Sec.~\ref{sec:disc} we conclude with a
discussion.

\section{Background}
\label{sec:background}

It has been proposed that the stretched exponential behavior exhibited
by the long time relaxation of $F({\bf q},t)$ can be attributed to a
sum of many independent local exponential relaxations with different
time constants, i.e., to a distribution of relaxation times
\cite{richter}. This interpretation is one form of the so-called
``heterogenous'' scenario for relaxation
\cite{doliwa,richter,ediger,spiess,hetero,heuer,sillescu}.  A number
of recent experiments~\cite{ediger,spiess,hetero} have shown that in
liquids such as orthoterphenyl and polystyrene within $10$~K of their
glass transition temperature $T_g$, subsets of molecules rotate slowly
relative to the rest of the molecules on time scales long compared
with collision times, but shorter than the relaxation time of density
fluctuations. These liquids were thus termed ``dynamically
heterogenous.'' None of these experiments were able to explicitly
demonstrate whether slow molecules are spatially correlated, but
typical distances over which slow molecules may be correlated were
inferred \cite{ediger}.

There have been numerous attempts to indirectly measure a
characteristic length scale over which molecular motions are
correlated at the glass transition both in
experiments~\cite{donth,mckenna,kremer,barut} and in
simulations~\cite{mountain95,ray}.  Donth~\cite{donth} relates
the distribution of relaxation times in systems approaching their
glass transition to equilibrium thermodynamic fluctuations having a
characteristic size of $\sim 3$ nm at $T_g$. Thermodynamic
measurements on orthoterphenyl~\cite{mckenna}, and dielectric
measurements on salol~\cite{kremer}, N-methyl-$\epsilon$-caprolactan
and propylene glycol~\cite{barut}, showed a shift in $T_g$ due to
confinement in pores of the order of a few
nanometers. Mountain~\cite{mountain95} showed that the size of regions
that support shear stress in a simulation of a glass-forming mixture
of soft spheres grows with decreasing temperatures.  Monte Carlo
simulations of polymer chains in two dimensions demonstrated strong
finite size effects on diffusion\cite{ray}. A number of experiments
and simulations on polymers confined to thin films all found a shift
of $T_g$ due to confinement \cite{jerome,wu,dutcher,mansfield,basch}.
These effects have all been attributed to the presence of
cooperatively rearranging regions that grow with decreasing $T$.
However, the origin of this characteristic length has never been shown
explicitly.  In particular, the connection of the characteristic
length to a cooperative mechanism of molecular motion has not been
experimentally demonstrated.

The intuitively-appealing picture of cooperative molecular motion was
proposed in 1965 by Adam and Gibbs~\cite{crc}. In their classic paper,
they proposed that significant molecular motion in a cold, dense fluid
can only occur if the molecules rearrange their positions in a
concerted, cooperative manner. They postulated that a glass-forming
liquid can be viewed as a collection of independently relaxing
subvolumes within which the motion of the particles is cooperative. As
the temperature of the liquid is lowered, the number of particles
involved in cooperative rearrangements increases. If structural
relaxation occurs through the cooperative rearrangement of groups of
molecules, the liquid observed over a time-scale shorter than the
structural relaxation time will appear as a collection of regions of
varying mobility. These predictions can be tested by selecting subsets
of molecules that relax slower (or faster) than the average, and
determining whether the molecules in a subset are randomly scattered
through the sample or tend to cluster in a characteristic way.

The explicit connection between dynamical heterogeneity and
cooperative motion is only recently being investigated experimentally
in detail\cite{kremer}.  However, there have been a number of recent
computational investigations addressing these issues.  For example,
Muranaka and Hiwatari \cite{hiwatari} showed that displacements of
particles measured over a timescale of the order of 5 collision times
are correlated within a range of about two interparticle distances in
a two-dimensional binary mixture of soft disks below the freezing
point. Wahnstr\"om \cite{wahnstrom} showed that hopping processes in a
strongly supercooled binary mixture are cooperative in nature.  Hurley
and Harrowell \cite{hurley} identified fluctuating local mobilities in
a supercooled two-dimensional (2-d) soft-disk system, and showed an
example of correlated particle motion on a timescale of the order of
20 collision times.  Mountain \cite{mountain95} demonstrated similar
correlated particle motion in a 2-d supercooled Lennard-Jones mixture.
By examining the time at which two neighboring particles move apart in
2-d and 3-d simulations of a supercooled soft-sphere mixture, Yamamoto
and Onuki demonstrated the growth of correlated regions of activity
\cite{onuki}. They further studied the effect of shear on these
regions \cite{onuki}, and showed that the size of the regions
diminished in high shear.  The clusters of ``broken bonds'' (denoting
pairs of neighboring particles that separate beyond the nearest
neighbor distance) identified in that work are similar in some 
respects to the clusters of highly mobile particles in a 3-d binary
Lennard-Jones liquid reported previously by us \cite{kdppg}, and
described in detail in the present paper.  The connection between the
clusters of Ref.~\cite{kdppg}, which demonstrate a form of dynamical
heterogeneity, and cooperative particle motion, was shown in
Ref.~\cite{ddkppg}.

\section{Simulation Details}
\label{sec:model}

We performed equilibrium molecular dynamics (MD) simulations of a
binary mixture (80:20) of $N=8000$ particles in three dimensions.  The
simulations were performed using the LAMMPS molecular dynamics code
\cite{lammps} which was designed for use on distributed memory
parallel machines.  LAMMPS partitions particles (atoms or molecules)
across processors via a spatial decomposition \cite{spatdecomp}
whereby each processor temporarily ``owns'' particles in a small fixed
region of the simulation box.  Each processor computes the motion of
its particles and exchanges information with neighboring processors to
compute forces and allow particles to migrate to new processors as
needed.

The 6400 particles of type A and 1600 particles of type B interact via
a 6-12 Lennard-Jones potential,
\begin{equation}
V_{\alpha \beta}(r) = 4 \epsilon_{\alpha \beta} \left[ \left( \frac{\sigma_{\alpha \beta}}{r}\right)^{12} 
- \left( \frac{\sigma_{\alpha \beta}}{r}\right)^6 \right],
\end{equation}
where $\alpha \beta \in \{A,B\}$.  The interaction forces between
particles are zero for all $r > r_c=2.5\sigma_{AA}$. Both types of
particles are taken to have the same mass $m$. The Lennard-Jones
interaction parameters $\epsilon_{\alpha,\beta}$ and
$\sigma_{\alpha,\beta}$ for this mixture are: $\epsilon_{AA}=1.0$,
$\epsilon_{AB}=1.5$, $\epsilon_{BB}=0.5$, $\sigma_{AA}=1.0$,
$\sigma_{AB}=0.8$, $\sigma_{BB}=0.88$.  Both the relative
concentration of particle types and the interaction parameters were
chosen to prevent demixing and crystallization
\cite{kobandersen}. Throughout this paper, lengths are defined in
units of $\sigma_{AA}$, temperature $T$ in units of
$\epsilon_{AA}/k_B$, and time $t$ in units of
$\sqrt{\sigma_{AA}^2m/\epsilon_{AA}}$.

The simulations for each state point ($P,T,\rho$) are performed in
three stages. First, a constant NPT adjustment run is performed by
coupling the system to stochastic heat and pressure baths to bring the
system from a nearby state point (usually the previously simulated
state point) to the desired state point~\cite{npt}.  Second, a
constant NVT equilibration run is performed to test for unwanted
drifts in pressure $P$ or potential energy $U$\cite{nvt}. If no drift
is observed, the final state of the system is considered to represent
an equilibrium state of the system.  Third, a constant NVE
data-gathering run is performed using the final equilibrated state
obtained from the second stage, and snapshots containing the particle
coordinates and velocities are taken at logarithmic time intervals
during the run. In this stage, the equations of motion are integrated
using the velocity Verlet algorithm with a step size of 0.0015 at the
highest temperature, and 0.003 at all other temperatures.  All
quantities presented here are calculated in this third stage. The
analysis is performed by post-processing the snapshot files, which
number as many as several thousand for the lower temperatures.

We simulated nine state points along a path in ${P,T,\rho}$ that is
linear when projected in the $(P,T)$ plane. This path was chosen so
that we would approach, from high temperature, the mode-coupling
critical point $T_c=0.435, P_c=3.03, \rho_c=1.2$
~\cite{kobandersen} along a path different from that used in
Ref.~\cite{kobandersen}. Table~1 shows the values of $P,T,$ and
$\rho$ for each state point.

For state points far above $T_c$ (e.g. runs 1--5), the data-gathering
runs required more cpu time than the equilibration.  For state points
nearer $T_c$, the equilibration stage was the most time consuming.  In
these cases the NVT stage of the simulations showed a slight drift of
the pressure over time. To shorten the time required for complete
equilibration, we estimated a new volume or temperature to create a
nearby state point that we expected would be very nearly equilibrated
with the current particle configuration.  Then we instantaneously
scaled the positions or velocities of the system (ie. adjusted the
volume or temperature) and began another constant NVT run to test for
equilibration.  By iterating this procedure a few times we were able
to find an equilibrated state point within $0.03\%$ of the desired $P$
and $T$ at low temperatures.

At the lowest $T$ studied ($T=0.4510$), the total run time following
equilibration is $1.2 \times 10^4$ time units. Thus, assuming Argon
values for the parameters in Eq.~3.1, the data presented here extend
up to 25.8 ns.

\section{Structure and Bulk Relaxation}
\label{sec:structure}

In this section, we show that the simulated liquid exhibits the
characteristic features of an atomic glass-forming
liquid.

\begin{figure}
\hbox to\hsize{\epsfxsize=1.0\hsize\hfil\epsfbox{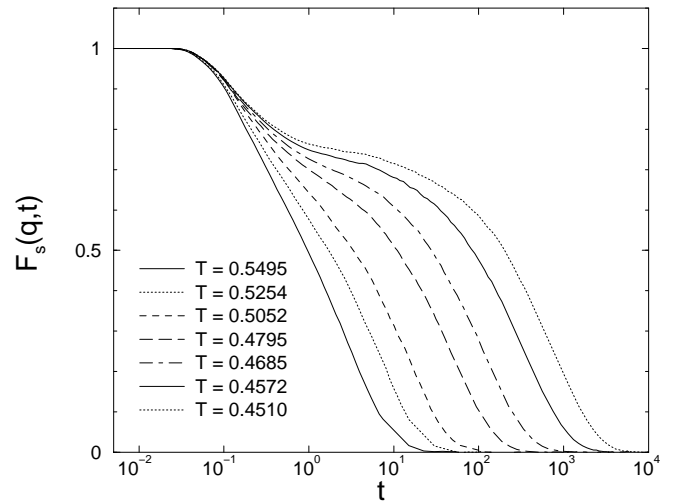}\hfil}
\caption{Incoherent (self) intermediate scattering function $F_s(q,t)$
for $q_{max}=7.12$.}
\label{figfqt}
\end{figure}

Structural relaxation may be probed experimentally by the intermediate
scattering function $F({\bf q},t)$, which is both the spatial Fourier
transform of the van Hove correlation function $G({\bf r},t)$ and the
inverse time transform of the dynamic structure factor $S({\bf
q},\omega)$ \cite{hansen}.  In a computer simulation, the self
(incoherent) part of the intermediate scattering function $F_s({\bf
q},t)$ may be calculated directly from
\begin{equation}
F_s({\bf q},t)= \frac{1}{N_{\alpha}}\Big\langle \sum_{j \in \alpha}
e^{i{\bf q} \cdot ({\bf r}_j(t) - {\bf r}_j(0))} \Big\rangle ,
\label{fqtv}
\end{equation}
where $r_j(t)$ is the position of particle $j$ at time $t$, and
$\langle \cdots \rangle$ indicates an average over independent
configurations. This quantity describes the relaxation of density
fluctuations due to single particle displacements on an inverse length
scale $2\pi/q$, where $q \equiv |{\bf q}|$.  If we assume rotational
invariance of the system, $F_s(q,t)$ depends only on $q$.  The time
dependence of $F_s(q,t)$ for the A particles for $q=q_{max}$ is shown
in Fig.~\ref{figfqt}.  (Throughout this paper, $q$ is chosen as
$q_{max}$, the position of the first maximum of the static structure
factor $S(q,0)$).  At high $T$, $F_s(q,t)$ decays to zero
exponentially.  As the system is cooled, $F_s(q,t)$ develops a plateau
that separates a short time relaxation process from a long time
relaxation process. This plateau indicates a transient
``localization'' of particles in the ``cages'' formed by their
neighbors, and is a characteristic feature of all glassforming
liquids.

The mode-coupling theory developed for supercooled liquids by G\"otze
and Sj\"ogren makes a number of predictions concerning the decay of
the intermediate scattering function \cite{mct}.  These predictions
have been tested and verified for the LJ potential used here in a
regime of $P$, $T$, and $\rho$ similar but not identical to that
simulated here \cite{kobandersen}.  There it was shown, e.g., that the
early and late $\beta$-relaxation regimes are well described by power
laws, and that the late time behavior of $F_s(q,t)$ exhibits
time-temperature superposition with a time constant $\tau_{\alpha}$
that diverges as a power law as $T$ approaches $T_c \simeq 0.432$,
with exponent $\gamma \simeq 2.7$. The diffusion constant was found to
scale as $D\sim(T-T_c)^{-\gamma}$, with $\gamma=2.0$ for the A
particles, $\gamma=1.7$ for the B particles, and $T_c = 0.435$.

 The simulations performed in the present work extend from a point in
the phase diagram where two-step relaxation begins to emerge, down to
a state point that is within approximately $4\%$ of $T_c$. Over this
range, we find that $\tau_{\alpha}$ increases by 2.4 orders of
magnitude, and fits well to the power law form found in
Ref.~\cite{kobandersen}, with approximately the same critical
temperature and critical exponent.

It is well known that although relaxation becomes strongly
nonexponential and relaxation times increase by many orders of
magnitude as a supercooled liquid approaches a glass transition,
changes in the static structure of most liquids are far less
remarkable.  To demonstrate this for our system, we examine the pair
correlation functions $g_{\alpha \beta}(r)$ given by
\begin{equation}
g_{\alpha \beta}({\bf r}) = \frac{V}{N_\alpha N_\beta}
\Big \langle \sum_{{i\in \alpha} \atop {j\in \beta}}  
\delta({\bf r} + {\bf r}_j -{\bf r}_i)\Big\rangle,
\label{grm}
\end{equation}
for $\alpha\neq\beta$ and
\begin{equation}
g_{\alpha \alpha}({\bf r}) = \frac{V}{N_\alpha (N_\alpha-1)}
\Big \langle \sum_{i,j\in \alpha}  
\delta({\bf r} + {\bf r}_j -{\bf r}_i)\Big\rangle,
\label{grm2}
\end{equation}
where $N_{\alpha}$ ($N_{\beta}$) is the total number of particles of
species $\alpha$ ($\beta$).  With this normalization,
$g_{\alpha\beta}({\bf r})$ converges to unity for $r \to \infty$ in
the absence of long range correlations. Assuming rotational
invariance, the correlation functions do not depend on the direction
of the vector ${\bf r}$, but only on the distance $r=|{\bf r}|$.

\begin{figure}
\hbox to\hsize{\epsfxsize=1.0\hsize\hfil\epsfbox{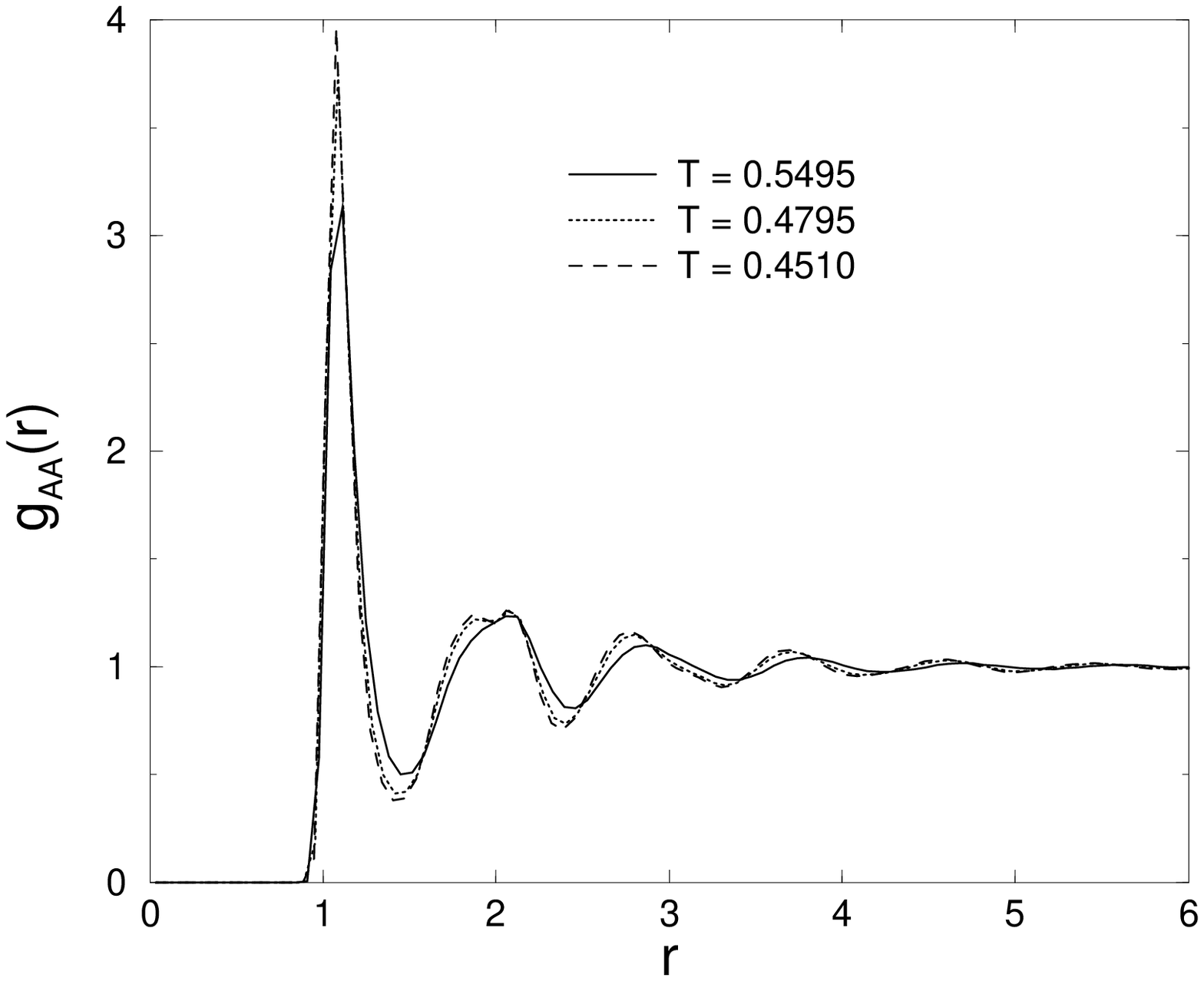}\hfil}
\caption{Pair correlation function $g_{AA}(r)$ of the A particles for three different temperatures.}
\label{figgr}
\end{figure}

\begin{figure}
\hbox to\hsize{\epsfxsize=1.0\hsize\hfil\epsfbox{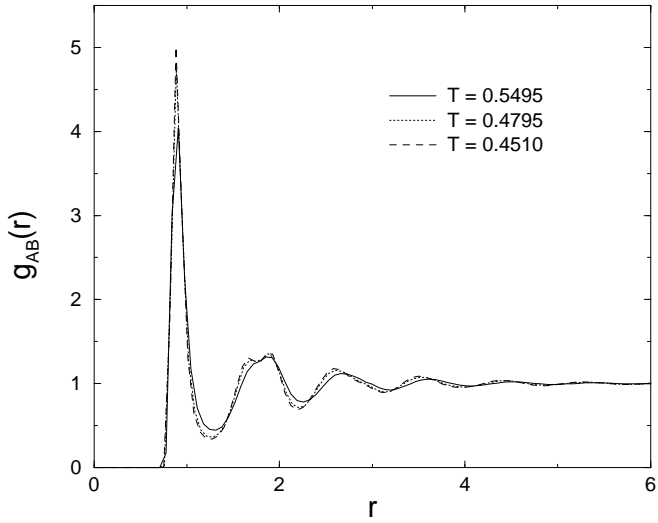}\hfil}
\caption{Pair correlation function $g_{AB}(r)$ between A and B particles for three different temperatures.}
\label{figgrab}
\end{figure}

\begin{figure}
\hbox to\hsize{\epsfxsize=1.0\hsize\hfil\epsfbox{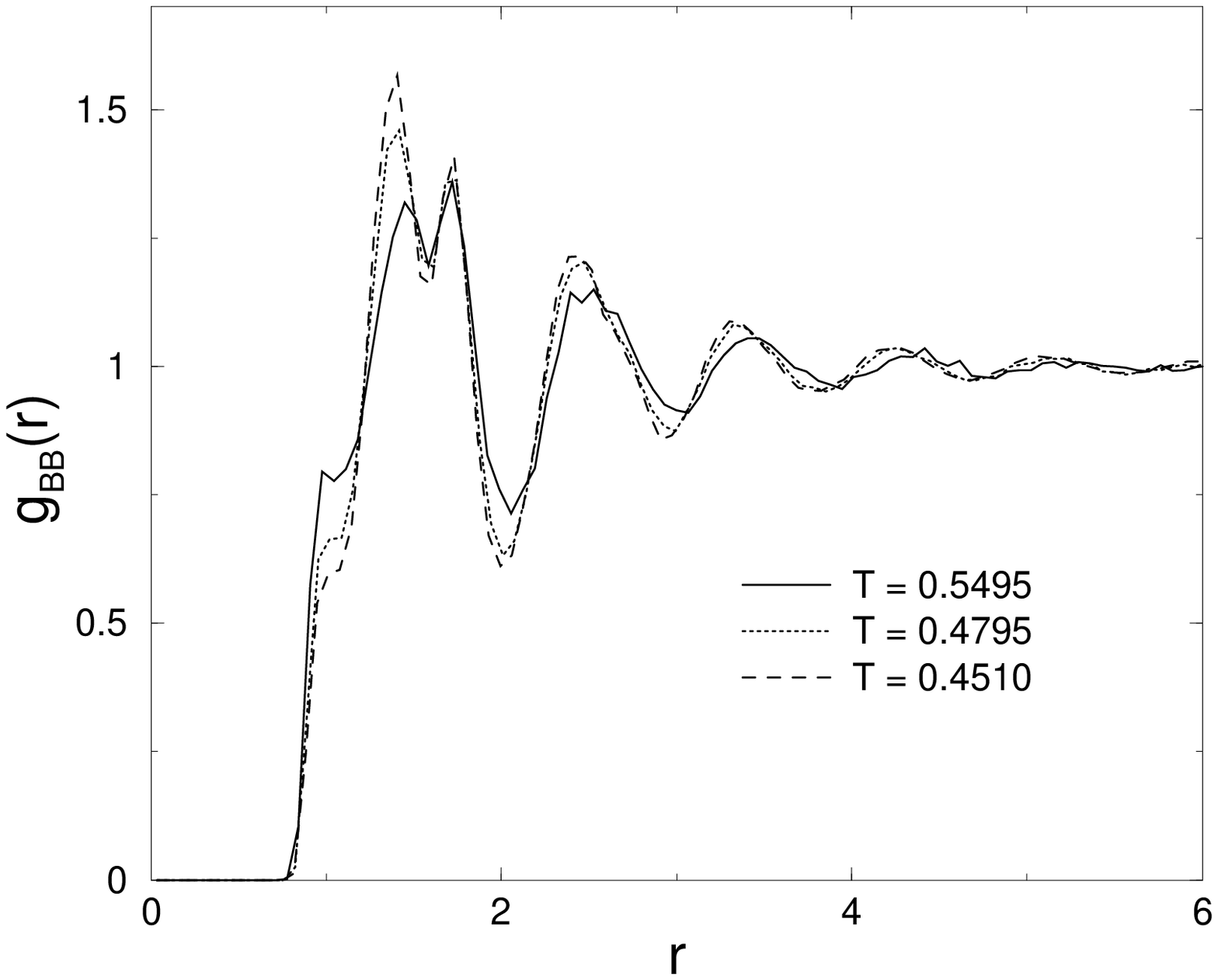}\hfil}
\caption{Pair correlation function $g_{BB}(r)$ of the B particles for three different temperatures.}
\label{figgrbb}
\end{figure}

In Figs.~\ref{figgr},\ref{figgrab} and \ref{figgrbb} we show the pair
correlation functions $g_{AA}(r)$, $g_{AB}(r)$, and $g_{BB}(r)$ for
three temperatures.  The figures show that these functions do not
change dramatically as a function of the state point.  As the
temperature is lowered, the main effect on all three functions is that
the maxima and the minima become slightly more pronounced.
Additionally, the second maximum of $g_{AA}(r)$ and $g_{AB}(r)$ at low
$T$ shows a splitting that has commonly been interpreted as a
signature of an amorphous solid, although at these state points our
system is an equilibrium liquid.  Recently, evidence has been
reported~\cite{truskett} that in a 2-d system of hard-disks the
splitting of the second peak in the pair correlation function is due
to the formation of regions with hexagonal close-packed order.

\section{Single Particle Dynamics}
\label{sec:dynamics}

Having established that the model liquid studied here exhibits the
characteristic bulk phenomena of a glass-forming liquid, we examine in
this section the distribution of individual particle motions.

\begin{figure}
\hbox to\hsize{\epsfxsize=1.0\hsize\hfil\epsfbox{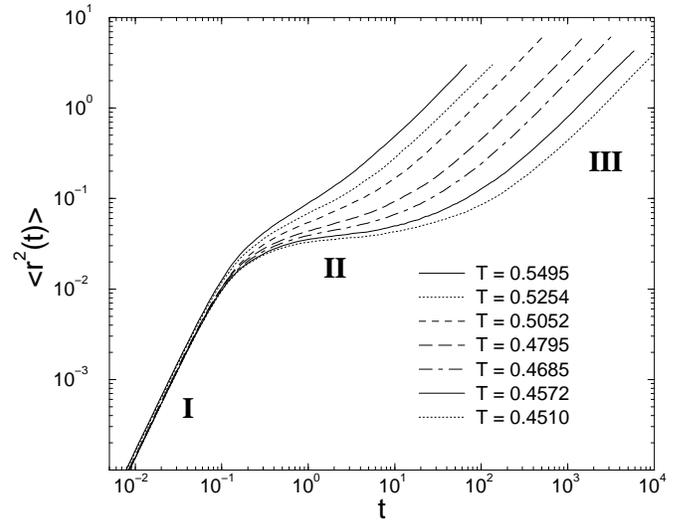}\hfil}
\caption{Mean square displacement $\langle r^2(t)\rangle$ of the A
particles vs. time.}
\label{r2oft}
\end{figure}

The most basic dynamical bulk quantity that is easily accessible to
simulation is the particle mean square displacement (MSD), $\langle
r^2(t)\rangle$.  Because we are investigating a binary mixture, we
refer in the following to a MSD for the A particles and a MSD for the
B particles.  At high $T$, the MSD for both species exhibits two
distinct regimes (see Fig. \ref{r2oft}).  In the short time limit
(regime I) the MSD is ballistic, i.e. $\langle r^2(t)\rangle \propto
t^2$. For longer times (regime III), the MSD is diffusive, i.e.
$\langle r^2(t)\rangle \propto t$.  As the system is cooled, an
intermediate regime (II) between these two limiting behaviors
develops.  Before entering the diffusive regime, $\langle
r^2(t)\rangle$ exhibits a plateau, analogous to the plateau in the
intermediate scattering function, that likewise arises from a
transient ``caging'' of each particle by its neighbors.  As seen in
the figure, the time the system spends in the plateau depends strongly
on $T$, and increases with decreasing $T$.  The MSD for the B
particles (not shown) exhibits qualitatively the same time dependence
as shown in Fig.~5 , but the diffusive regime is reached at shorter
times, and the diffusion constant is larger, than for the A particles
\cite{kobandersen}.  This difference can be explained by the different
sizes of the A and B particles and by the fact that the interaction
constant $\epsilon_{BB}$ is smaller than $\epsilon_{AA}$.

\begin{figure}
\hbox to\hsize{\epsfxsize=1.0\hsize\hfil\epsfbox{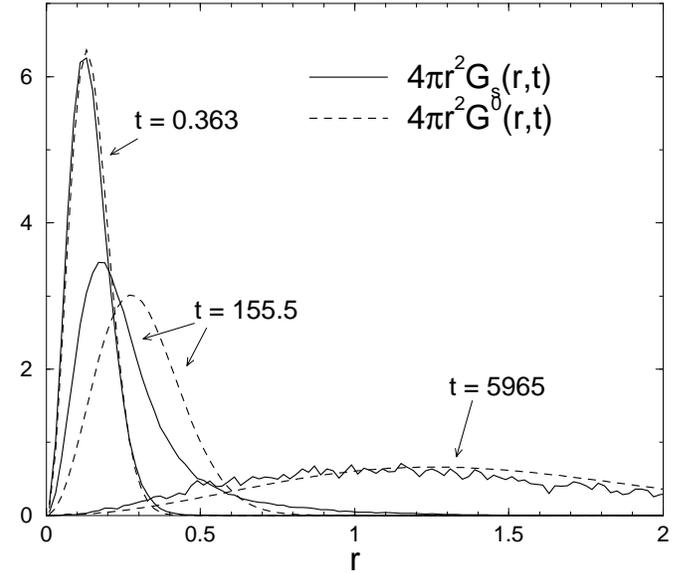}\hfil}
\caption{Solid line: $4 \pi r^2G_s(r,t)$ of the A particles 
for three times at $T=0.4510$. 
Dashed line: Gaussian approximation calculated using the measured
$\langle r^2(t)\rangle$ for the same three times.  } 
\label{fig1}
\end{figure}

\begin{figure}
\hbox to\hsize{\epsfxsize=1.0\hsize\hfil\epsfbox{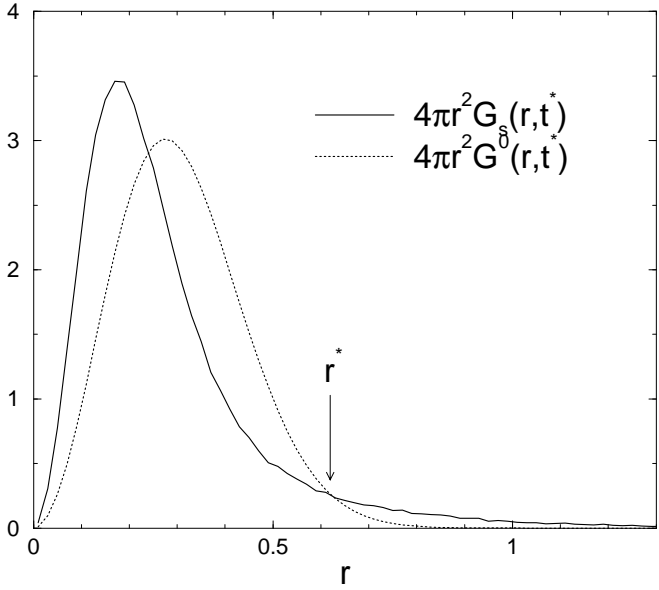}\hfil}
\caption{Same intermediate time data as in previous figure, but
enlarged. Solid line: $4 \pi r^2G_s(r,t)$ of the A particles at
$t=155.5$ at $T=0.4510$.  Dashed line: Gaussian approximation
calculated using the measured $\langle r^2(t)\rangle$ for the same
time.}
\label{fig:figtail}
\end{figure}

In this paper, we are interested in whether spatial correlations exist
between particles that exhibit either extremely large or extremely
small displacements over some time interval.  To determine this, we
must first define the time interval over which the particle
displacements will be monitored.  Obviously, displacements may be
monitored over any time interval, from the ballistic regime to the
diffusive regime.  To see whether there is a natural time scale on
which the particle displacements might exhibit a particularly strong
correlation, we turn to the self part of the van Hove correlation
function, $G_s(r,t)$, which gives the probability to find a particle
at time $t$ at a distance $r$ from its position at $t=0$
\cite{hansen}:
\begin{equation}
G_s({\bf r},t)=\frac{1}{N_\alpha}\Big\langle 
\sum_i \delta({\bf r}+{\bf r}_i(0)-{\bf r}_i(t))\Big\rangle.
\label{gis}
\end{equation}
Due to the rotational symmetry of the system, $G_s({\bf r},t)$ is a 
function of the modulus $r$ of the vector displacement ${\bf r}$.  The 
quantity $4 \pi r^2 G_s(r,t)$, which gives the number of particles located 
a distance $r$ from their original position at time $t$, is shown in 
Fig.~\ref{fig1} for the A particles for three different times at the lowest 
$T$.
Also shown in Fig.~\ref{fig1} is the Gaussian approximation $4\pi r^2
G^0(r,t)$ , where \cite{hansen}
\begin{equation}
G^0(r,t)=\left(\frac{3}{2 \pi \langle 
r^2(t)\rangle}\right)^{\frac{3}{2}}
\exp \left( \frac{3 r^2}{2\langle r^2(t)\rangle} \right)
\end{equation}
and where $\langle r^2(t) \rangle$ is equal to the measured one.  The
Gaussian form appears to be a good approximation to $G_s(r,t)$ at both
short and long times.  However, it is apparent from the figure that
$G_s(r,t)$ is significantly different from $G^0(r,t)$ at intermediate
times.  In particular, while many of the particles have traveled less
than would be expected from the knowledge of $\langle r^2(t) \rangle $
alone, a small number of particles have traveled significantly
farther.  As a result, at intermediate times $G_s(r,t)$ displays a
long tail that extends beyond one interparticle distance
at $T=0.4510$ (cf. Fig.~\ref{fig:figtail}).

\begin{figure} 
\hbox to\hsize{\epsfxsize=1.0\hsize\hfil\epsfbox{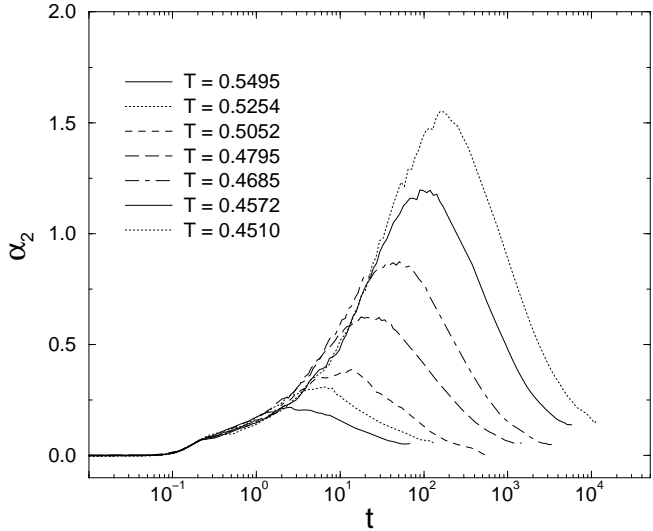}\hfil}
\caption{Non-Gaussian parameter
$\alpha_2(t)$ vs. time for different temperatures.}
\label{figalpha} 
\end{figure}

This ``long tail'' behavior is most pronounced at a time $t^*$ when
$G_s(r,t)$ deviates most from a Gaussian (cf. Fig.~\ref{fig:figtail})
as characterized by the ``non-Gaussian'' parameter\cite{rahman},
\begin{equation}
\alpha_2(t)=\frac{3 \langle r^4(t)\rangle}{5 \langle r^2(t)\rangle^2}-1, \ \
\ \ d=3.
\end{equation}
While one can define many different non-Gaussian parameters, this
particular one involves the lowest possible moments.  With this
definition, $\alpha_2(t)$ is zero if $G_s(r,t)$ is Gaussian. If a
distribution has a tail that extends to distances exceeding those for
a Gaussian distribution with the same second moment, all higher order
moments of this distribution will exceed those of the corresponding
Gaussian, and consequently $\alpha_2(t)$ will assume positive
values. In Fig.~\ref{figalpha} we show $\alpha_2(t)$ for various $T$
for the A particles. As expected, $\alpha_2(t)$ is zero at short
times, then becomes positive, exhibits a maximum, and finally goes to
zero at long times. As $T$ decreases, the position of the maximum
$t^*$ shifts towards longer times, and the height of the maximum
$\alpha_2^*$ increases.  For all $T$, we find that $t^*$ corresponds
to times in the late-$\beta$/early-$\alpha$ relaxation
regime. Furthermore, by dividing $\alpha_2(t)$ by $\alpha_2^*$, and
dividing $t$ by $t^*$, one can show \cite{alpha} that all curves
collapse onto a single master curve {\it for all times larger than the
microscopic time} (where they already collapse before scaling)
\cite{hurley,kob,fs-alpha}. This data collapse, which is likely related to
the time-temperature superposition exhibited by the intermediate
scattering function (but not trivially related, since $t^*$ does not
scale linearly with $\tau_{\alpha}$), suggests that $t^*$ is in some
sense a characteristic time for this system \cite{alpha}. Note that
$t^*$ is orders of magnitude larger than the microscopic ``collision
time'' \cite{collision} $\tau$; for example, at $T=0.4510$ $t^*=155.5$
and $\tau=0.09$.
  
We see from an analysis of the temperature-dependent distribution of
particle displacements at various times, and the calculation of the
time-dependence of the non-Gaussian parameter, that the single
particle dynamics is most non-Gaussian --- and displays the widest
range of possible behaviors --- on the timescale $t^*$.  The interval
from zero to $t^*$ thus provides a convenient choice over which to
monitor the particle displacements and study their correlations
because (i) since $t^*$ is the time at which the distribution of
particle displacements is broadest, it may also be when the liquid is
likely to be most ``dynamically heterogeneous''; and (ii) $t^*$ is
well-defined and easily calculated.  Thus, throughout this paper we
will use the time window from zero to $t^*$ as the time interval over
which the particle displacements are calculated, and over which we
investigate dynamical heterogeneity. 

\section{Analysis of Spatial Correlations of Particle Displacements}
\label{sec:mobility}

In this paper, we are interested in studying the extreme behavior of
the individual particle motion, from the extremely mobile to the
extremely immobile.  In Refs.~\cite{kdppg} and \cite{ddkppg}, we
defined the magnitude of the displacement $u_i(t,t^*) \equiv |{\bf
r}_i(t^*+t)-{\bf r}_i(t)|$ of particle $i$ in a time interval $t^*$,
starting from its position at an arbitrarily chosen time origin $t$,
as a measure of the mobility of the $i$-th particle.  At $t^*$, the
distribution of $u_i$ values is given by the self part of the van Hove
correlation function, $G_s(r,t^*)$, where $r \equiv u$
(cf. Fig.~\ref{fig:figtail}).  In Refs.~\cite{kdppg,ddkppg} a subset
of ``mobile'' particles was defined by selecting all the particles
that in the interval $[0,t^*]$ had traveled beyond the distance $r^*$
where $G_s(r,t^*)$ exceeds $G^0(r,t^*)$.  With this definition,
``mobile'' particles are those that contribute to the long tail of the
van Hove distribution function at the time $t^*$
(cf. Fig.~\ref{fig:figtail}).  In Refs.~\cite{kdppg,ddkppg}, it was
shown that mobile particles selected according to this rule tend to
cluster \cite{kdppg}, and move cooperatively \cite{ddkppg}. This
definition of mobility given by the magnitude of particle displacement
is thus sufficient to establish the phenomena of both dynamical
heterogeneity and cooperative motion.

Intuitively, we think of immobile particles as those particles which
are trapped in cages formed by their neighbors. Nevertheless,
particles do not sit at one position; they essentially ``oscillate''
back and forth within the cage formed by their neighbors.  To study
correlations between the most immobile particles, we need a definition
of mobility which allows us to select the particles for which the
amplitude of this oscillation (ie. the maximum displacement of the
particle) is the smallest.  In this paper, we therefore define the mobility
$\mu_i(t)$ of the i-th A particle as the {\it maximum} distance reached
by that particle in the time interval $[t,t+t^*]$:
\begin{equation}
\mu_i(t)={\rm max}_{t'\in[0,t^*]} \{|{\bf r}_i(t'+t)-{\bf r}_i(t)|\}
\label{mu}
\end{equation}
This new definition of mobility, which we use throughout this paper,
allows us to examine different subsets of particles, from the most to
the least mobile, in the same way. As a compromise between examining
the most extreme behavior and including enough particles to obtain
good statistics when examining their spatial correlation
(i.e. maximizing the signal-to-noise ratio), we will examine the 5\%
most mobile and 5\% least mobile particles. {\it Thus we define as
``mobile'' the 5\% of all particles having the highest values of
$\mu(t)$, and ``immobile'' the 5\% having the lowest value.}  Note
that this new definition of mobility does not qualitatively change the
results obtained previously in~\cite{kdppg,ddkppg}, provided that the
new definition selects approximately the same fraction of the sample
as the definition previously used (approximately 5.5\% in
Ref.~\cite{kdppg}) Compare, for instance, Figs.~\ref{figgrm} and
\ref{figratio} with Fig.~3 of Ref.~\cite{kdppg}.

\begin{figure}
\hbox to\hsize{\epsfxsize=1.0\hsize\hfil\epsfbox{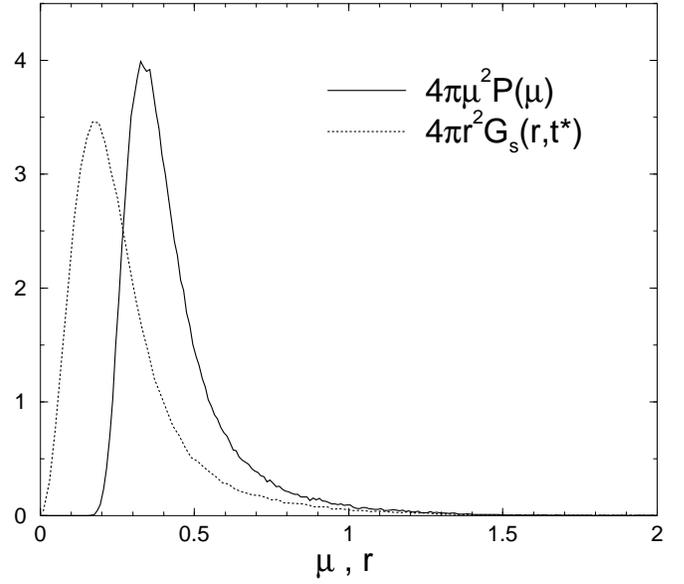}\hfil} 
\caption{Probability distribution $4\pi \mu^2 P(\mu,t^*)$ (dashed line) of a
particle having a maximum displacement of magnitude $\mu$ at
$t^*$. For comparison, the distribution $4 \pi r^2 G_s(r,t^*)$ is also shown.}
\label{fig:maxdisp}
\end{figure}

The subsets of mobile particles selected using the definition of
Ref.~\cite{kdppg} and that used here have a large overlap, since
particles that have moved relatively far at some time in the interval
$[0,t^*]$ are likely to remain relatively far at the end of the
interval. However, subsets of immobile particles selected with the two
different rules do not have as large an overlap, since a particle with
a small displacement at some time may have previously traveled far,
and then returned to its original position.  The distribution $4 \pi
\mu^2 P(\mu,t^*)$ at $t^*$ is shown in Fig.~\ref{fig:maxdisp}. For
comparison, the probability distribution $4 \pi r^2 G_s(r,t^*)$ is
also shown. Note that, although at $t^*$ particles can be found
arbitrarily close to their position at $t=0$, $P(\mu,t^*)$ is zero for
$\mu < 0.17$.

\begin{figure}
\hbox to\hsize{\epsfxsize=1.0\hsize\hfil\epsfbox{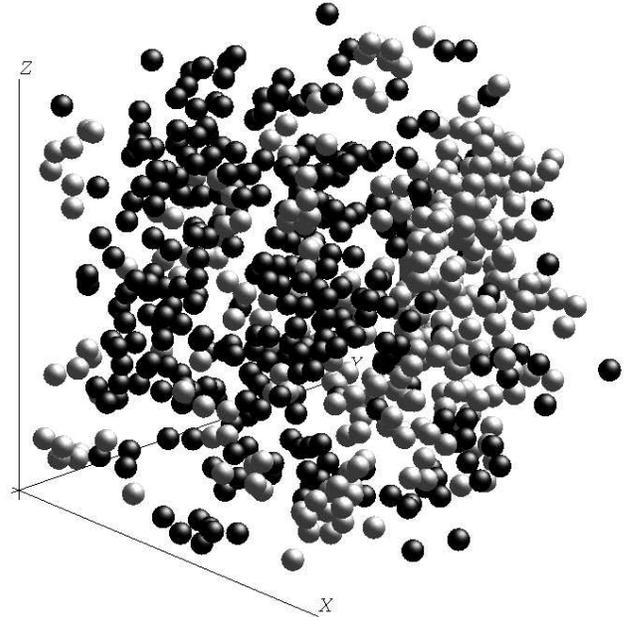}\hfil} 
\caption{The 320 mobile particles (light spheres) and the 320 immobile
particles (dark spheres) in a configuration at an arbitrarily chosen
time.}
\label{figslice-m}
\end{figure}

\begin{figure}
\hbox to\hsize{\epsfxsize=1.0\hsize\hfil\epsfbox{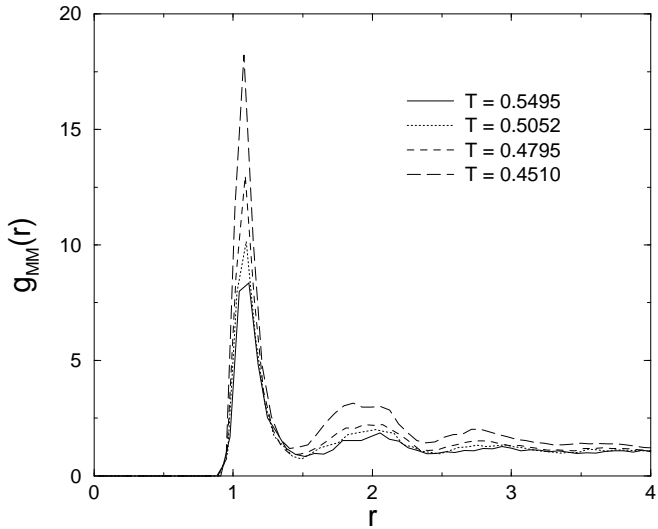}\hfil} 
\caption{Pair correlation function $g_{MM}(r)$ between mobile A
particles at four different temperatures.}
\label{figgrm}
\end{figure}

\begin{figure} 
\hbox to\hsize{\epsfxsize=1.0\hsize\hfil\epsfbox{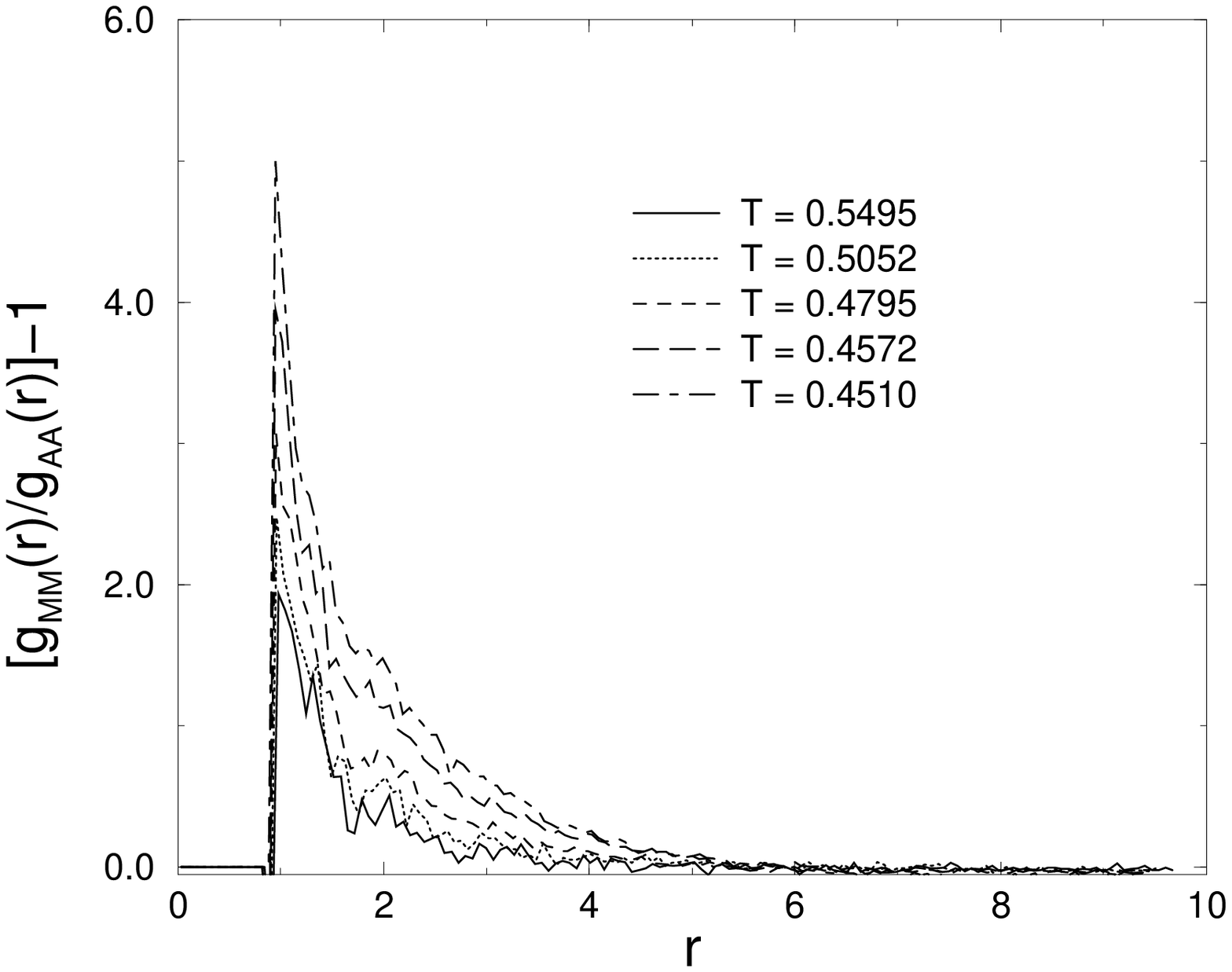}\hfil}
\caption{$\Gamma(r)=[g_{MM}(r)/g_{AA}(r)] -1$ vs. $r$ for different 
temperatures.}
\label{figratio}
\end{figure}

In Fig.~\ref{figslice-m}, we show the 320 mobile particles (light
spheres) and the 320 immobile particles (dark spheres) at the
beginning of an arbitrary time interval $[t,t+t^*]$ for one 
configuration at $T=0.4510$. The other 7360 particles are not
shown. The figure shows that particles of similar mobility are
spatially correlated and that particles with different mobility tend
to be anticorrelated. These correlations can be quantitatively studied by
calculating static pair correlation functions between particles
belonging to the different subsets.

In Fig.~\ref{figgrm} we show the pair correlation function $g_{MM}(r)$
between mobile particles for four different temperatures.  $g_{MM}(r)$
is defined by Eq.~\ref{grm2} with the sum restricted to the mobile
particles..  For all $T$, $g_{MM}(r)$ is appreciably higher that the
average $g_{AA}(r)$ (cf. Fig.~\ref{figgr}) for all $r$. The ``excess''
correlation given by the ratio $\Gamma(r)=[g_{MM}(r)/g_{AA}(r)] - 1$
is plotted as a function of $r$ in Fig.~\ref{figratio}.  With the
exception of the excluded volume sphere of the LJ potential,
$\Gamma(r)>0$ at intermediate distances and converges to zero for
large $r$. It is clear from the figure that the total excess
correlation, given by the area under the curve, increases with
decreasing $T$.

\begin{figure}
\hbox to\hsize{\epsfxsize=1.0\hsize\hfil\epsfbox{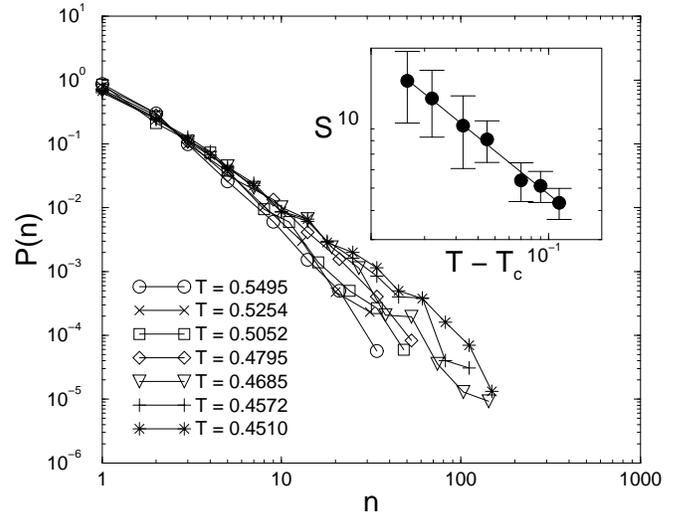}\hfil}
\caption{Distribution of the size $n$ of clusters of mobile
particles. Inset: Mean cluster size $S$ plotted versus $T-T_c$, where
is the fitted MCT critical temperature $T_c=0.435$. The straight line
is a power law fit $S \sim (T-T_c)^{-\gamma}$, with $\gamma=0.618$.}
\label{clustersize}
\end{figure}

We can obtain an estimate of the typical distance over which mobile
particles are correlated by identifying clusters of nearest-neighbor
mobile particles~\cite{clusters}. To do this, we use the following
rule: two particles belong to the same cluster if their distance at
$t=0$ is less than $r_{nn}$, the radius of the nearest neighbor shell,
which is defined by the first minimum in $g_{AA}(r)$ and has a weak
temperature dependence.  In our hottest run $r_{nn}=1.45$, while in
the coldest run $r_{nn}=1.40$.  The distribution $P(n)$ of clusters of
size $n$ is shown in Fig.~\ref{clustersize}. Although most of the
clusters have only a modest size, the data show that a significant
fraction of the mobile particles, which themselves make up only 5\% of
the sample (320 particles), are part of big clusters. For instance, at
$T=0.4510$, there is typically at least one cluster in each
configuration that contains $\approx 100$ particles. For that $T$,
$P(n) \sim n^{-\tau}$ with $\tau = 1.86$.  In the inset we show the
mean cluster size $S = \sum n^2P(n) / \sum n P(n)$~\cite{esse},
plotted log-log versus $T-T_c$, where $T_c=0.435$ is the fitted
critical temperature of the mode coupling
theory~\cite{kobandersen,kob}. Although there is less than a decade on
either axis, the figure shows that the temperature dependence of $S$
is consistent with a divergence at $T_c$ of the form $S \sim
(T-T_c)^{-\gamma}$, with $\gamma \approx 0.618$.  Note that MCT makes
no predictions about clustering or the divergence of any length scales
as the critical point is approached \cite{lengthscale}.

\begin{figure}
\hbox to\hsize{\epsfxsize=1.0\hsize\hfil\epsfbox{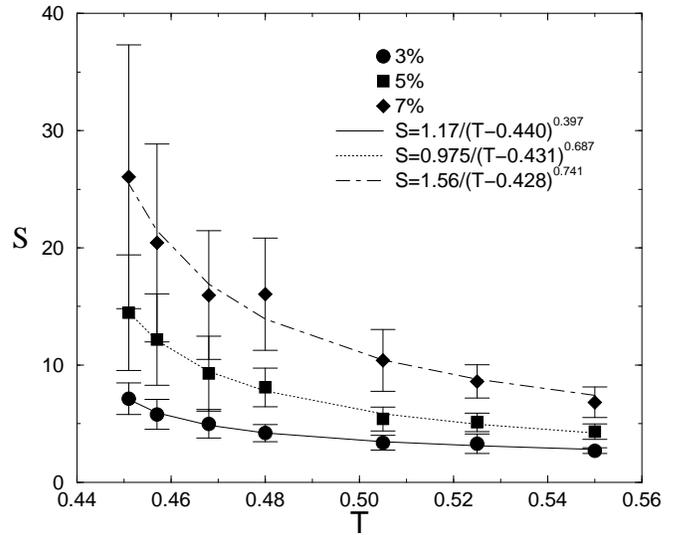}\hfil}
\caption{Mean cluster size $S$ plotted versus $T$, for subsets
containing $3\%, 5\%$ and $7\%$ of the most mobile particles. The data
for the $5\%$ are the same as those shown in the inset of the previous
figure.  The lines are power law fits $S \sim (T-T_p)^{-\gamma}$.
Best fit parameters are $T_p=0.440$, $0.431$ and $ 0.428$,
respectively, and $\gamma=0.397$, $0.687$, and $0.741$, respectively.}
\label{figsensitivity}
\end{figure}

To test the sensitivity of the apparent percolation transition at the
mode-coupling temperature, we repeat the cluster size distribution
analysis for the 3\% and 7\% most mobile particles. For each subset,
the mean cluster size $S$ is shown vs. $T-T_c$ in
Fig.~\ref{figsensitivity}. The best fit of $S \sim (T-T_p)^{-\gamma}$
to each set of data gives $T_p = 0.440$ for the set containing the 3\%
most mobile particles, $T_p=0.431$ for the set containing the $5\%$
most mobile particles, and $T_p = 0.428$ for the set containing the
7\% most mobile particles. However, within the accuracy of the data
the three sets are also consistent with a divergence at $T_c$.  If we
further increase the fraction of mobile particles beyond the fraction
corresponding to a random close-packed percolation transition
\cite{random}, the mobile particles percolate and most of the mobile
particles are found in a single cluster that spans the whole
simulation box.

\begin{figure}
\hbox to\hsize{\epsfxsize=1.0\hsize\hfil\epsfbox{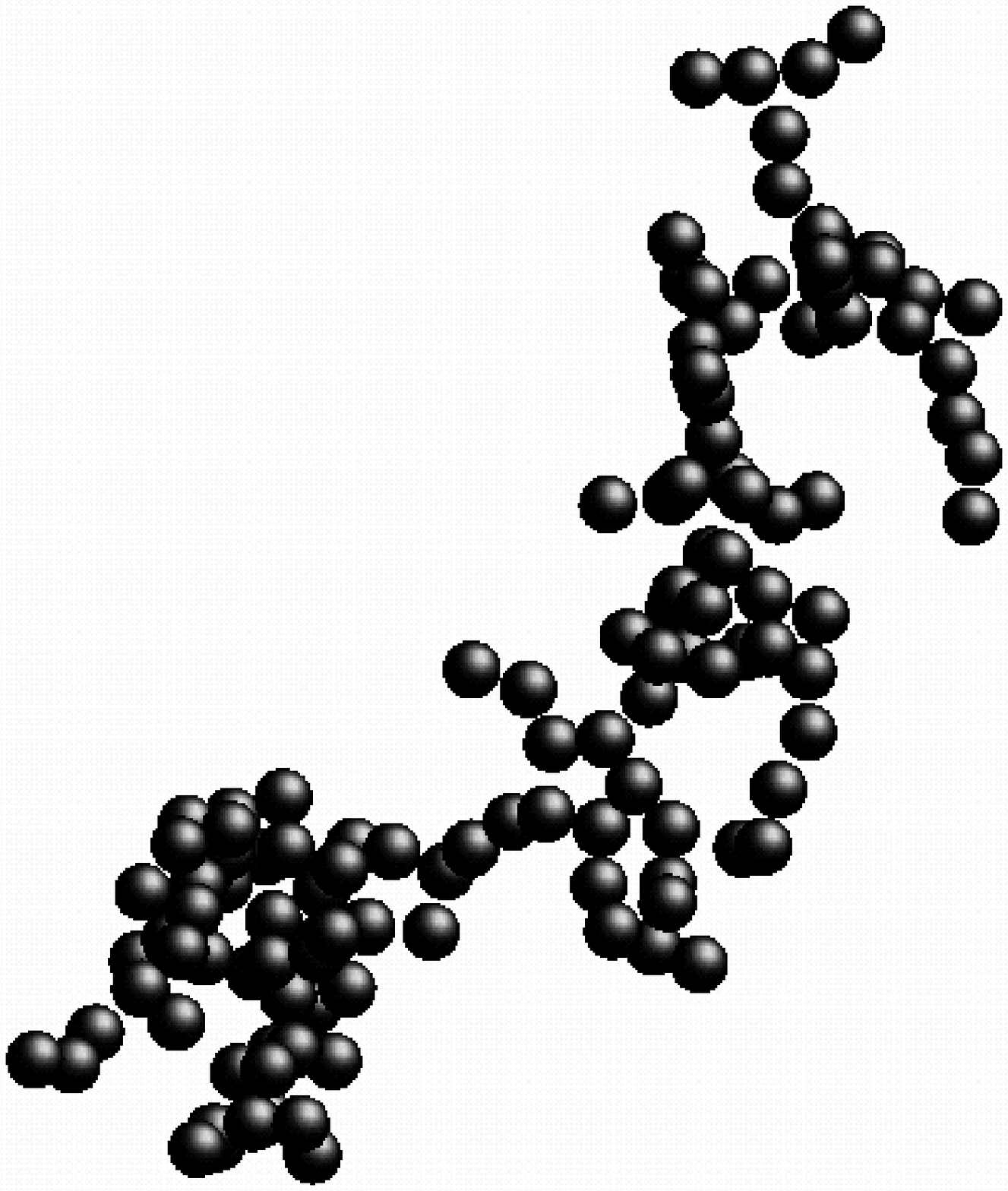}\hfil}
\caption{One of the largest clusters of mobile A particles found at
T=0.4510. The cluster is composed of 125 particles, which are
represented here as spheres of radius $r= 0.5 \sigma_{aa}$.}
\label{mcluster}
\end{figure}

In Fig.~\ref{mcluster} we show one of the largest clusters of mobile
particles found in our coldest simulation.  It is evident from the
figure that these clusters cannot be described as compact, as often
supposed either implicitly or explicitly in phenomenological models of
dynamically heterogeneous liquids~\cite{ediger,stillinger}. Instead,
the clusters formed by the mobile particles appear to have a disperse,
string-like nature.  As discussed in \cite{ddkppg}, a preliminary
calculation of the fractal dimension of the clusters, although
hampered by a lack of statistics, indicates that the clusters
have a fractal dimension close to 1.75, similar to that for both
self-avoiding random walks and the backbone of a random percolation
cluster in three dimensions \cite{stauffer}.

\begin{figure}
\hbox to\hsize{\epsfxsize=1.0\hsize\hfil\epsfbox{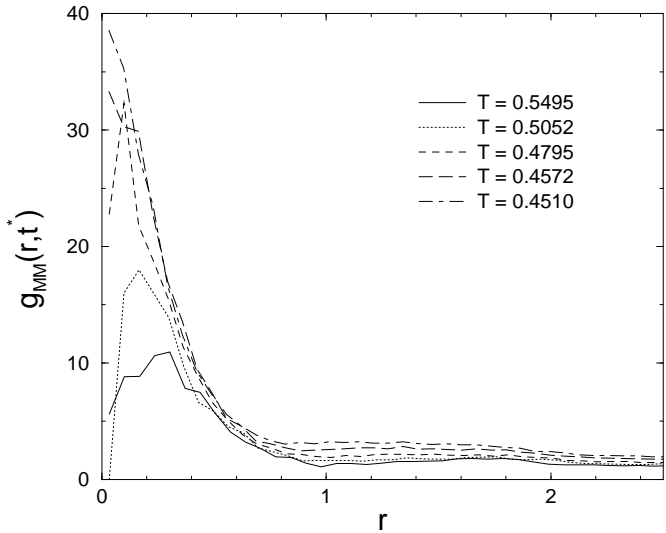}\hfil}
\caption{Time-dependent pair correlation function $g_{MM}(r,t^*)$
vs. $r$, for different temperatures.}
\label{figfollow}
\end{figure}

In Ref.~\cite{ddkppg}, it was shown that this quasi-one-dimensionality
appears to arise from the tendency for mobile particles to follow one
another.  This is demonstrated in Fig.~\ref{figfollow}, where we plot
the time-dependent pair correlation function for the mobile particles,
$g_{MM}(r,t^*)$ for different temperatures. At $t=0$, this function
coincides with $g_{MM}(r)$ in Fig.~\ref{figgrm}. For $t >0$, the
nearest neighbor peak moves toward $r=0$, demonstrating that a mobile
particle that at $t=0$ is a nearest neighbor of another mobile
particle tends to move toward that particle at later times. We find
that the peak at $r=0$ is highest near $t = t^*$, and decreases for
later times.  A small but discernable peak at $r=0$ is also present in
$g(r,t^*)$ \cite{strings}.

Fig.~\ref{fig-string} shows a cluster of mobile particles
at two different times, $t=0$ and $t=t^*$, to demonstrate the
cooperative, string-like nature of the particle motion. 

In a manner identical to our analysis of the mobile particles, we
define as immobile the $5\%$ of the A particles that have the lowest
value of $\mu$. The pair correlation function $g_{II}(r)$ between
immobile particles shown in Fig.~\ref{grimm} shows that these
particles also tend to be spatially correlated.  It is interesting to
note that while the maxima in $g_{II}(r)$ are higher at all $T$ than
the corresponding maxima in $g_{AA}(r)$, the depth of the minima does
not change appreciably for the lowest temperatures. 
Fig.~\ref{ratiogrimm} shows the ratio $\Gamma(r) = [g_{II}(r)/g_{AA}(r)] -1$
as a function of $r$.  In contrast to what we find for the most
mobile particles, the correlation between immobile particles does not
show any evidence of singular behavior as $T$ decreases.  Instead, the
correlation appears to grow and then ``saturate'' to some limiting
behavior for all $T < 0.468$. Moreover, Fig.~\ref{ratiogrimm} shows
that the local structure of the liquid appears
to be more ordered in the vicinity of an {\it immobile} A particle
than in the vicinity of a {\it mobile} A particle.

In Fig.~\ref{clustersize-imm} we show the size distribution of the
clusters of immobile particles, formed with the same rule used for the
mobile ones.  One of the largest clusters found at $T=0.4510$ is shown
in Fig.~\ref{icluster}. In the inset of Fig.~\ref{clustersize-imm} we
show the mean cluster size $S$ versus $T-T_c$.  We find that the mean
cluster size of immobile particles is relatively constant with
$T$. This may be because immobile particles are relatively
well-packed, and cannot grow beyond some limiting size
\cite{kivelson}. Or, these clusters may be the ``cores'' of larger
clusters of particles with small displacements, that may grow with
decreasing $T$.  To elucidate this, more particles (e.g. the next 5\%
higher mobility) should be included in the analysis.  We will return
to this important point and provide further relevant data in the next
section.

The correlation between mobile and immobile particles, measured by the
pair correlation function $g_{MI}(r)$ (Fig.~\ref{grmobimm}), shows
that mobile and immobile particles are anti-correlated.  A comparison
between $g_{MI}(r)$ and $g_{AA}(r)$, shown in
Fig.~\ref{ratio-imm-mob}, demonstrates that, over several
interparticle distances, the probability to find an immobile A
particle in the vicinity of a mobile one is lower than the probability
to find a generic A particle. The figure also shows that the
characteristic length scale of the anticorrelation grows with
decreasing $T$. This length scale does not show a tendency to diverge
as $T_c$ is approached. In particular, the curves for the two coldest
runs (and closest to $T_c$) are almost coincident.

\section{Local Energy and Local Composition vs. Mobility}
\label{sec:energy}

We have seen in the previous section that despite the lack of a
growing static correlation, a growing {\it dynamical} correlation ---
characterizing spatial correlations between particles of similar
mobility --- does exist.  These correlations must therefore arise from
subtle changes in the local environment that are not completely
captured by the usual static pair correlation function.  In this
Section, we calculate several quantities to elucidate whether the
mobility of a particle is related to its potential energy, and to the
composition of its local neighborhood.

In Fig.~\ref{figene} we show the distributions of the potential
energies of the 5\% most mobile, 5\% least mobile, and all particles
at $T=0.4510$, calculated at the beginning of an arbitrary time
interval $[t,t+t^*]$. The distributions have been normalized such that
the area under each curve is one. The distributions differ by a small
relative shift of the mean value, approximately 3\% for the high
mobility distributions and somewhat less for the low mobility
distribution. We find that the magnitude of the shift increases with
decreasing $T$, but the {\it relative} shift appears to be {\it
independent} of $T$. Since the liquid is in equilibrium, this shift
will vanish for $t \to \infty$. Thus, not suprisingly, mobile
particles are those that in a time $t^*$ are able to rearrange their
position so as to lower their potential energy.  It is worth noting
that the mobility does not show any correlation with the kinetic
energy of the particles measured at $t=0$.  The kinetic energy
distributions of the subsets with different mobility coincide exactly
with the average distribution, showing that the mobility cannot be
related to the presence of ``hot spots'' in the liquid.

We next divide the entire population of A particles into 20 subsets,
each composed of $5\%$ of the particles.  In the first subset we put
the $5\%$ of the particles with the highest values of $\mu$ (the
mobile particles defined above), in the second subset the next $5\%$,
and so on.  The last subset thus contains the $5\%$ most immobile
particles.  In Fig.~\ref{figdistrene} we plot (on the x-axis) the
average mobility of each subset versus (on the y-axis) the average
potential energy of that subset at $t=0$.  We find that the subset
with the lowest mobility is also the one with the lowest potential
energy.  We also find that as the potential energy increases, the
mobility increases.  We see from the figure that the mobile particles
are the subset with the highest average potential energy at $t=0$.

Two more points are worth noting in Fig.~\ref{figdistrene}.  First, at
all $T$ the mobile particles move, on average, approximately one
interparticle distance in the time interval $[0,t^*]$.  Second, for
all $T$ the difference in both mobility and potential energy between
the $5\%$ most mobile particles and the next subset is significantly
larger than between any other two consecutive subsets. This
observation suggests that the choice of 5\%, while arbitrary, is a
reasonable one.  As shown in the figure, the separation between the
$5\%$ most mobile particles and the next subset shows a tendency to
grow with decreasing $T$.  Note however, that the distance between the
lowest mobility subset and the next subset {\it decreases} with
decreasing $T$, making it very difficult in the current approach to
define an appropriate subset containing particles whose mobility is
distinctly lower than the rest.  This, together with the result that
the mean cluster size of immobile particles is relatively constant
over the range of temperatures studied, suggests that our analysis of
the lowest subset is inadequate to fully characterize clusters of
particles which do not move a substantial distance \cite{immobile}.

Thus we see that the gross structural information contained in the
potential energy is sufficient to establish a general correlation
between energy and mobility. However, as seen in Fig.~\ref{figene},
the distribution of potential energies of mobile particles overlaps
for most of the range of the abscissa with the distribution for the
generic A particles.  {\it Thus, it is not possible to decide if a
certain particle is mobile on the basis of that particle's
instantaneous potential energy alone.}  Other factors, such as defects
in the local packing, and the relative potential energy of neighboring
particles, must contribute as well.

The relation between mobility and potential energy suggests a relation
between mobility and local composition.  Indeed, a calculation of the
pair correlation functions $g_{MA}(r)$ and $g_{MB}(r)$ (see
Figs.~\ref{figgma} - \ref{figgmbgab}) between a mobile particle and a
generic A or B particle, respectively, shows that, on average, a
mobile A particle tends to have less B particles, and more A
particles, in its nearest neighbor shell than a generic A particle.

A correlation can also be found between {\it immobility} and small
composition fluctuations of the mixture.  A comparison of $g_{AB}(r)$
to the pair correlation function $g_{IB}(r)$, which measures the
number of B particles a distance $r$ from a test immobile A particle,
shows that an immobile A particle has, on average, more B particles in
its nearest neighbor shell than does a generic A particle.  As shown
in Fig.~\ref{ratio-imm-b}, where the ratio $[g_{IB}/g_{AB}(r)]-1$ is
plotted as a function of $r$, this enhanced correlation is independent
of $T$, and therefore does not suggest any evidence of $A-B$ phase
separation (recall that the chosen energy parameters preclude $A-B$
phase separation).

From these results it is clear how a correlation between mobility and
local composition causes a correlation between mobility and potential
energy.  Since the attractive interaction between $A$ and $B$
particles is stronger than either the attractive $AA$ or $BB$
interaction, the presence of a $B$ between two $A$'s reduces their
potential energy.  $A$ particles in a $B$-rich region can thus be
expected to have a reduced mobility.  $A$ particles in a $B$-poor
region, however, will have a higher potential energy, resulting in a
higher mobility.  

\section{Structural Relaxation of Particle Subsets and Dynamical Heterogeneity}
\label{sec:lifetime}

We have shown that it is possible to select subsets of particles
according to their maximum displacement over a timescale in the region
of the late $\beta$-early $\alpha$ relaxation. We have also shown that
the particles belonging to subsets selected at the extrema of the
mobility spectrum are spatially correlated, and are related to small
fluctuations in the local potential energy, and, consequently, in the
local composition of the mixture.  All of the data presented here
suggest that this supercooled liquid contains fluctuations in local
mobility, with diffuse, quasi-one-dimensional regions of high
mobility, and relatively compact regions of low mobility.

To measure how long a mobile particle will continue to be mobile, we
define a variable $\nu_i^M(t)$ as $1$ if the $i$-th particle belongs
to the subset of the $5\%$ most mobile particles in the interval
$[t,t+t^*]$, and $0$ otherwise. The function $\sigma_M(t)$
\begin{equation}
\sigma_M(t)=\frac{1}{n_M-\frac{n_M^2}{N_A}} \Big(\sum_i \langle
\nu_i^M(t)\nu_i^M(0)\rangle -\frac{n_M^2}{N_A}\Big),
\label{sigma}
\end{equation}
measures the fraction of particles that are mobile in the interval
$[0,t^*]$ and still mobile in the interval $[t,t+t^*]$, when the time
origin is shifted by $t$. Here $n_M$ is the number of mobile particles
(320 in the present case), and $N_A$ is the total number of A
particles (6400).  The second term on the right-hand side of
Eq.~\ref{sigma} is the number of particles that by random statistics
would be classified as mobile in both time intervals.  The
normalization of $\sigma_M(t)$ is chosen so that $\sigma_M(0)=1$. The
results for $\sigma_M(t)$ for the coldest $T$ are shown in
Fig.~\ref{figsigma}.

We have also measured the fraction of particles that are immobile in
the interval $[0,t^*]$ and still immobile in the interval
$[t,t+t^*]$. Analogous to the case for the mobile particles, we define
$\sigma_I(t)$ as
\begin{equation}
\sigma_I(t)=\frac{1}{n_I-\frac{n_I^2}{N_A}} \Big(\sum_i \langle
\nu_i^I(t)\nu_i^I(0) \rangle -\frac{n_I^2}{N_A}\Big),
\label{sigmai}
\end{equation}
where $n_I$ is the number of immobile particles (320) and $\nu_i^I(t)$
is a function that is $1$ if the i-th particle is an immobile one in
the interval $[t,t+t^*]$, and $0$ otherwise. The function
$\sigma_I(t)$ is also shown in Fig.~\ref{figsigma}.

The functions $\sigma_M(t)$ and $\sigma_I(t)$ are memory functions of
mobility.  When they have decayed to zero, there are no particles that
have retained memory of their mobility in the initial time
interval. Because a particle's mobility is based upon a criterion that
depends on $t^*$, certain time-dependent functions measured for these
subsets will have some ``kink'' at $t^*$. If a different $t^*$ is
chosen, the kink will move to the new $t^*$.  In this respect, there
is no ``natural'' lifetime for these clusters --- by definition, they
survive for a time $t^*$ \cite{gdp}. 

Nevertheless, we can obtain information from the form of the decay
both before and after $t^*$. Because the data was stored not less than
every 3 time units, we are unable to calculate $\sigma_M(t)$ and
$\sigma_I(t)$ for $t < 3$.  However, we see that these functions decay
substantially before this time, since already at $t=3$ both functions
are significantly smaller than one.  After this initial short-time
relaxation, a second decay of both functions is observed up to
$t=t^*$. At this time, a third decay process appears for the mobile
particles, and possibly also for the immobile particles. The main
point of Fig.~\ref{figsigma} is that beyond $t^*$, both functions are
less than 0.1.  Thus there is only a small tendency for particles to
retain memory of their mobility beyond the initial time interval
\cite{return}.

Thus, mobile and immobile regions do not persist beyond the time $t^*$
over which the particle mobility is monitored.  After the observation
time, mobile and immobile particles maintain little memory of their
previous state. Therefore the strong correlations found between
particles must arise from the motion itself. If, for instance, the
mobility of a particle and its spatial correlation with other
particles of similar mobility could be explained solely by local
fluctuations in quantities like density or composition, the mobility
should persist until these fluctuations decay to zero.  Instead, the
dependence of the lifetime on the observation time can be explained if
one assumes that particles can move only in a cooperative
manner. Indeed, as was shown in Ref.~\cite{ddkppg}, clusters of mobile
particles like that shown in Fig.~\ref{mcluster} can be decomposed
into numerous, smaller string-like clusters (``strings'') of particles
which follow one another in a cooperative fashion.  
 
Fig.~\ref{f_q_t-mob} shows the intermediate scattering functions
$F^M_s(q,t)$ and $F^I_s(q,t)$, defined as the spatial Fourier
transform of the self part of the van Hove correlation function
$G_s^M(r,t)$ or $G_s^I(r,t)$ of the mobile and immobile particles,
respectively. Both functions are identical to the bulk $F_s(q,t)$ for
times less than the ``collision time'' $\tau=0.09$ \cite{collision}.
The figure shows that a two-step relaxation process occurs for the
mobile particles, although the height of the plateau is smaller than
for the bulk. The presence of the plateau in $F^M_s(q,t)$ indicates that
the mobile particles are subject to the same ``cage effect''
experienced by the other particles, although the effective cage
``size'' and ``lifetime'' are different. Thus clusters of mobile
particles should not be thought of as ``fluidized'' regions of the
liquid in the simple sense that those regions might behave like high
temperature or low density liquids. Instead, the difference between
the mobile particles and the rest of the sample appears to be, from
the point of view of the single particle dynamics, that they
``escape'' the cage earlier than the other particles.

We also see that the three curves in Fig.~\ref{f_q_t-mob} cannot be
superimposed by scaling the time axis in the same way as one can
superimpose $F(q,t)$ curves for different temperatures.  Again,
this indicates that the mobile and immobile subsets are not simply
``hotter'' or ``colder'' subsets of the sample, in agreement with the
perfect superposition of the kinetic energy distributions.

In contrast to the bulk average $F_s(q,t)$, $F^M_s(q,t)$ is not a
monotonically decreasing function of time. For times longer than
$t^*$, a small but clearly detectable increase of the function can be
noticed in Fig.~\ref{f_q_t-mob}.  This behavior can be interpreted as
a tendency of a small fraction of the particles that we have selected
to return towards their position at the beginning of the selection
interval. These particles may also be those that contribute to the
small memory effect observed in $\sigma_M(t)$ in Fig.~\ref{figsigma},
but further analysis is required to establish this connection.

Finally, we show in Fig.~\ref{gateway} the fraction $\phi$ of
particles that at time $t$ have not yet been labeled mobile. This
function is calculated by labeling the mobile particles in the first
interval $[0,t^*]$, and then shifting the interval by $t$ and
reassigning the particle mobilities. Thus at $t=0$, 95\% of the
particles have not been labeled mobile. In the interval $[t,t+t^*]$,
more particles will have been labeled mobile, so $\phi$ will decrease.
We have normalized $\phi(t)$ such that $\phi(0) = 1$. Also shown in
Fig.~\ref{gateway} is the long-time $\alpha$-relaxation part of the bulk
$F_s(q,t)$ for $q=q_{max}$.  Fits to both functions are also shown.
Both functions fit well to a stretched exponential $y(t) =
A$exp$[(-t/\tau)^\beta]$ with $\beta = 0.75$ and $\tau_{\alpha} = 655$
for the intermediate scattering function, and with $\beta = 0.78$ and
$\tau_{\alpha} = 475$ for $\phi$. That both functions have a similar
form (similar $\beta$), and similar time constants, suggests that the
process by which immobile particles become mobile governs the long
time structural relaxation of density fluctuations at wavevectors
corresponding to the peak of the static structure factor.  Moreover,
it demonstrates that the ``arbitrary'' choice of 5\% represents a
physically meaningful fraction of the system .

\section{Discussion}
\label{sec:disc}

In this paper, we have described an investigation of the individual
particle dynamics of a cold, dense Lennard-Jones mixture well above
the glass transition in an effort to discover if the liquid is
dynamically heterogenous, and if so to determine the extent and nature
of the dynamical heterogeneity.  Since there were no quantitative
theoretical predictions regarding this matter, the approach we have
taken is exploratory; particle trajectories were saved during the
course of the simulation and then analyzed and visualized in numerous
ways. We find that this supercooled liquid is ``dynamically
heterogeneous'' because particles with similar mobility are spatially
correlated.  Note that our definition of heterogeneity is different
from the one used, for instance, in 4-D NMR experiments, where the
system is defined as heterogeneous if a slow subset remains slow for
times longer than the average relaxation time \cite{spiess}. We
further find that highly ramified clusters of mobile particles grow
with decreasing $T$ and appear to percolate at the mode-coupling
temperature. This is the first evidence for a percolation transition
coincident with $T_c$, and it is very different from the type of
percolation transition proposed in free volume theory \cite{cohen}.
It is especially interesting since MCT does not make any predictions
regarding clusters or diverging length scales.  We also find that
particles of low mobility form relatively well-ordered, compact
clusters which do not appear to grow with decreasing $T$ if the number
of immobile particles included in the subset is kept constant.
Although mobile and immobile clusters are anti-correlated, there is no
tendency towards bulk phase separation of mobile and immobile regions
because of the highly ramified, extended nature of the mobile regions.

In our analysis, we find no evidence to support a picture in which the
system can be thought of as a collection of subvolumes that each relax
independently and simultaneously with their own time
constant. Instead, it appears that at any given time, most particles
are localized in cages and a small percentage of particles form large
clusters of smaller, cooperatively rearranging ``strings.'' After
rearranging, these mobile particles become caged themselves, and
others become mobile. This process repeats until, on the time scale of
the $\alpha$ relaxation, each particle has rearranged at least once.
Thus the structural relaxation of the liquid appears to be highly
cooperative in the spirit of Adam and Gibbs, but where different
subvolumes of the liquid are able to relax only after other subvolumes
relax.  This will be further explored in a separate publication
\cite{allegrip}.

PHP acknowledges the support of NSERC. WK is partially supported by 
Deutsche Forschungsgemeinschaft under SFB~262.  CD and SCG thank Jack 
Douglas for many interesting discussions.

\bigskip
\noindent
{\it Corresponding author: sharon.glotzer@nist.gov.}

\bigskip
\bigskip
\bigskip
\begin{quote}
\begin{tabular}{|p{1cm}|p{2cm}|p{2cm}|p{2cm}|}
\hline
Run & $T$ & $P$ & $\rho$ \\
\hline
1 & 0.5495 & 0.4888 & 1.0859 \\
2 & 0.5254 & 1.0334 & 1.1177 \\
3 & 0.5052 & 1.4767 & 1.1397 \\
4 & 0.4899 & 1.8148 & 1.1553 \\
5 & 0.4795 & 2.0488 & 1.1651 \\
6 & 0.4737 & 2.1746 & 1.1705 \\
7 & 0.4685 & 2.2959 & 1.1757 \\
8 & 0.4572 & 2.5490 & 1.1856 \\
9 & 0.4510 & 2.6800 & 1.1910 \\
\hline
\end{tabular}

Table 1:Temperature T, pressure P  and density $\rho$ of the 
nine state points simulated.
\end{quote}
\bigskip

\bigskip
\noindent

\noindent

\noindent

\noindent

\noindent

\noindent

\noindent

\noindent

\noindent

\noindent

\noindent

\noindent

\noindent

\noindent

\noindent

\noindent

\noindent
\begin{figure}
\caption{A cluster of mobile particles at $t=0$ (light spheres) and
$t=t^*$ (dark spheres), for $T=0.4510$.}
\label{fig-string}
\end{figure}

\noindent
\begin{figure}
\hbox to\hsize{\epsfxsize=1.0\hsize\hfil\epsfbox{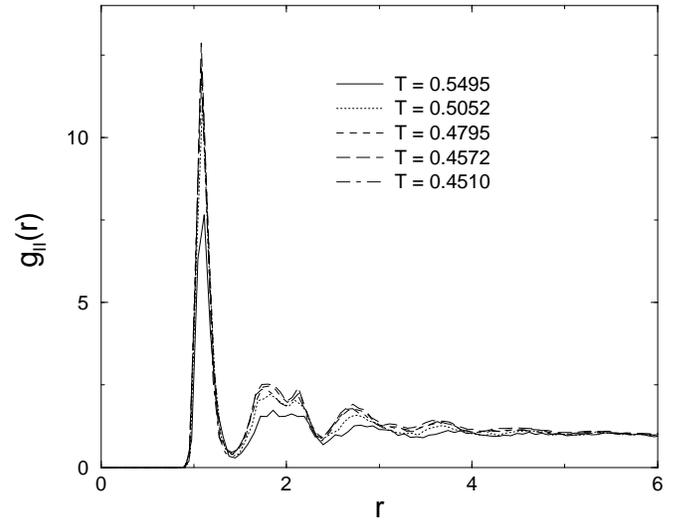}\hfil}
\caption{Pair correlation function $g_{II}(r)$ between immobile
particles.}
\label{grimm}
\end{figure}

\noindent
\begin{figure}
\hbox to\hsize{\epsfxsize=1.0\hsize\hfil\epsfbox{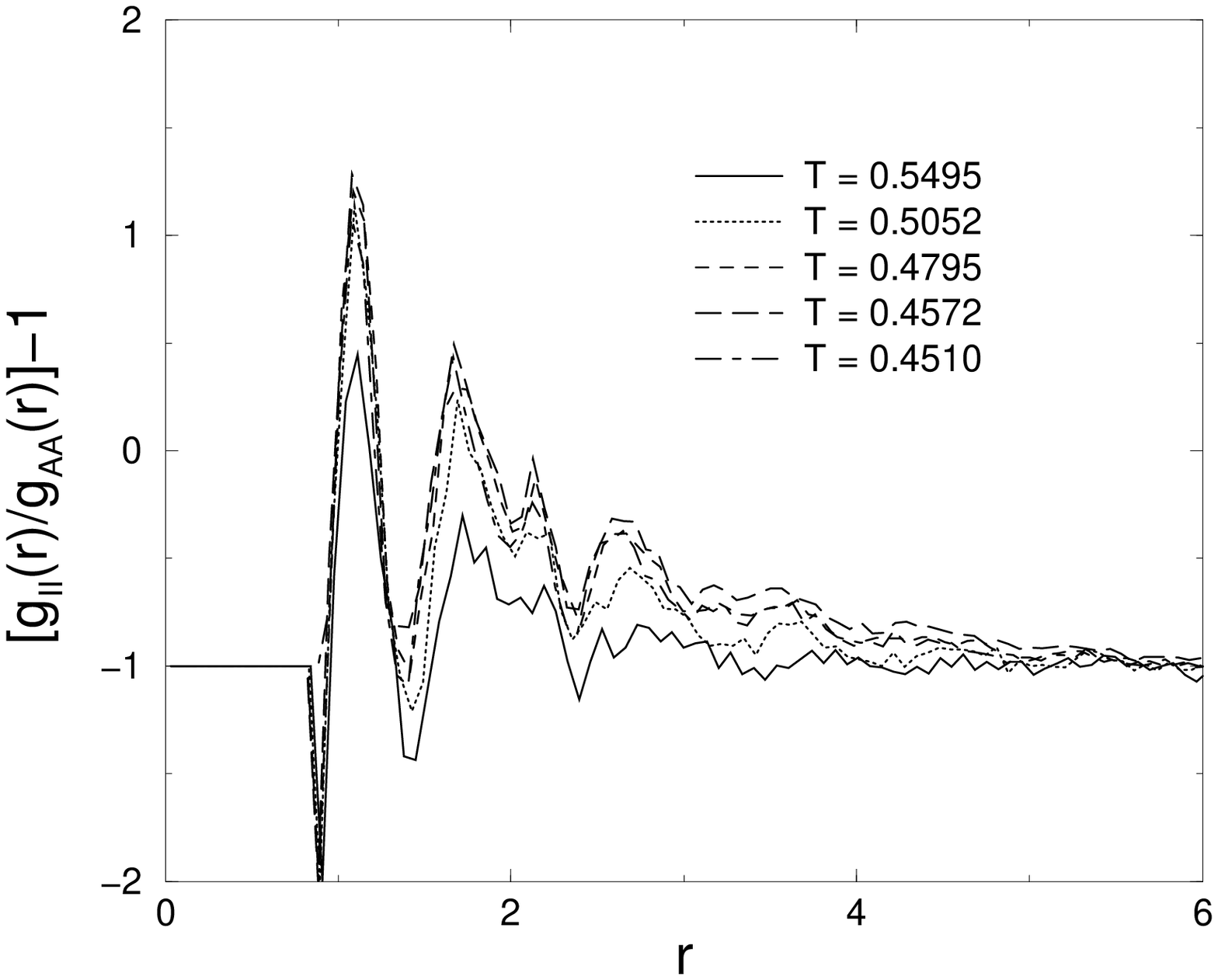}\hfil}
\caption{$\Gamma(r)=[g_{II}(r)/g_{AA}(r)] -1$ vs. $r$ for different 
temperatures.}
\label{ratiogrimm}
\end{figure}

\noindent
\begin{figure}
\hbox to\hsize{\epsfxsize=1.0\hsize\hfil\epsfbox{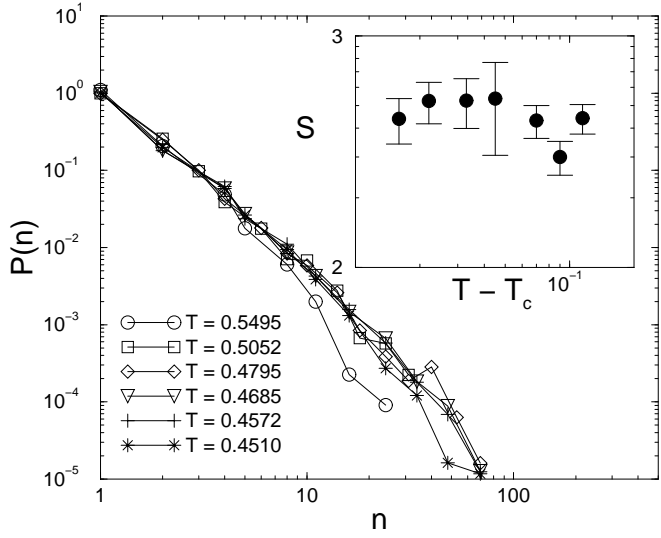}\hfil}
\caption{Distribution of the size $n$ of clusters of immobile
particles. Inset: mean cluster size $S$ plotted
vs.~$T-T_c$.}
\label{clustersize-imm}
\end{figure}

\noindent
\begin{figure}
\hbox to\hsize{\epsfxsize=1.0\hsize\hfil\epsfbox{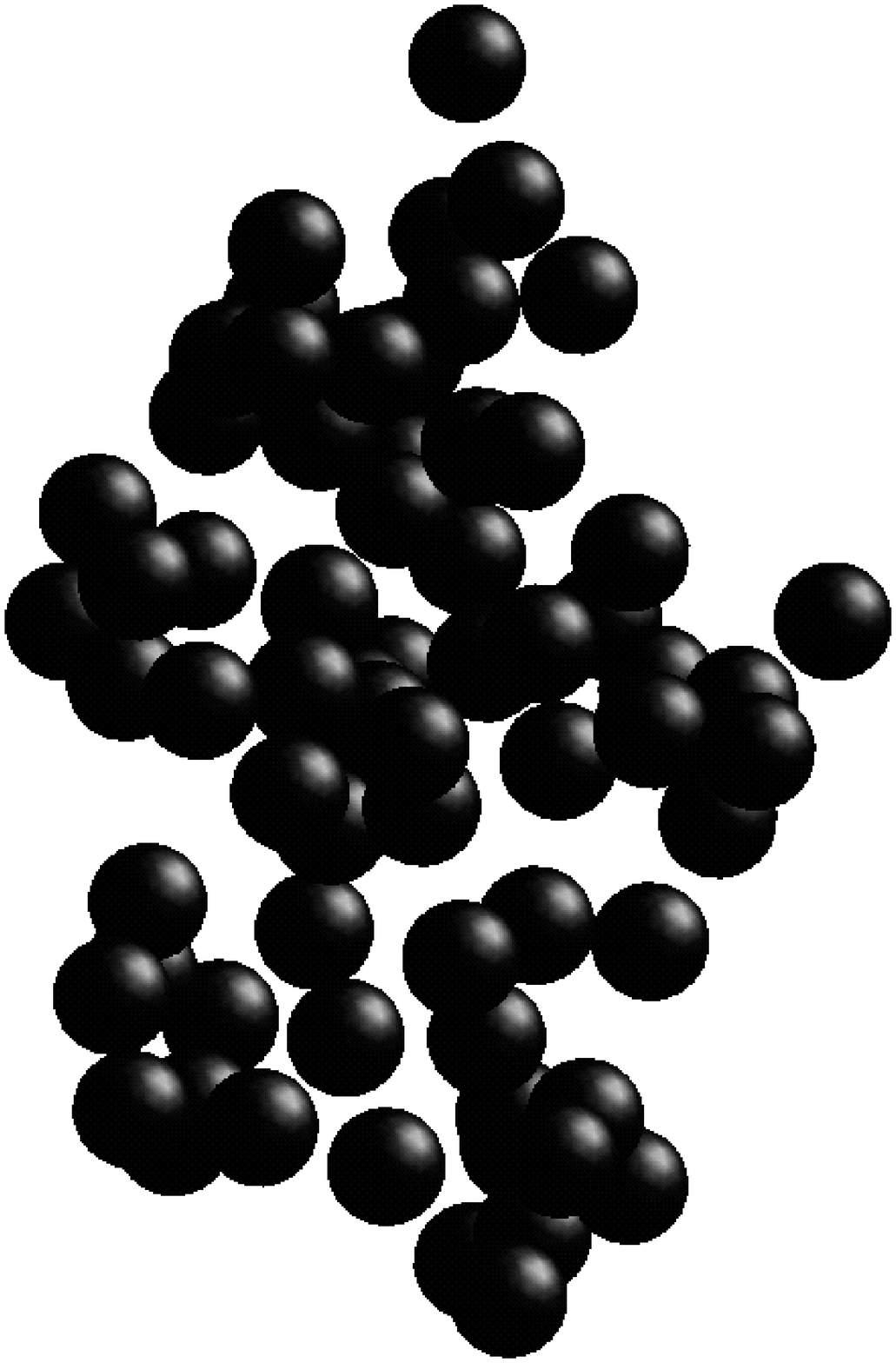}\hfil}
\caption{One of the largest clusters of immobile A particles found at
T=0.4510.  The cluster is composed of 70 particles, which are
represented here as spheres of radius $r= 0.5 \sigma_{aa}$}
\label{icluster}
\end{figure}

\noindent
\begin{figure}
\hbox to\hsize{\epsfxsize=1.0\hsize\hfil\epsfbox{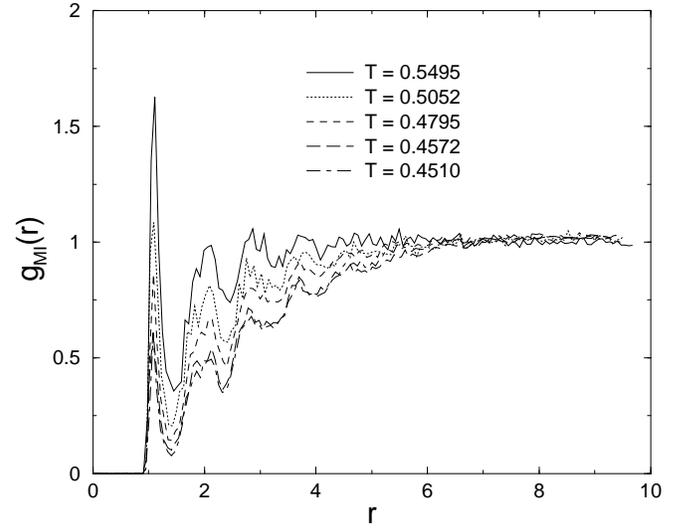}\hfil}
\caption{Pair correlation function $g_{MI}(r)$ between mobile 
and immobile A particles.}
\label{grmobimm}
\end{figure}

\noindent
\begin{figure}
\hbox to\hsize{\epsfxsize=1.0\hsize\hfil\epsfbox{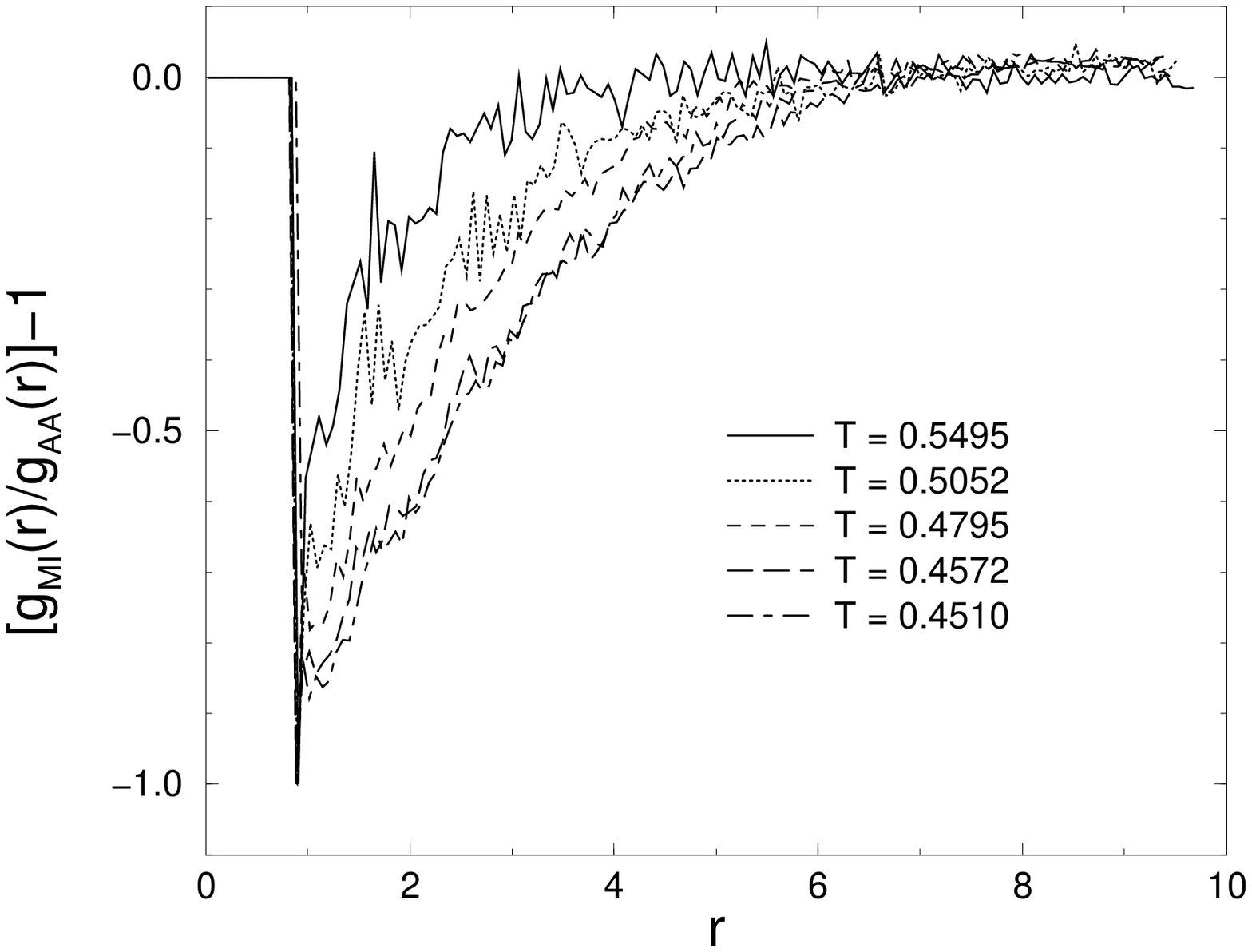}\hfil}
\caption{$\Gamma(r) = [g_{MI}(r)/g_{AA}(r)] - 1$ vs. $r$ for different 
temperatures.}
\label{ratio-imm-mob}
\end{figure}

\noindent
\begin{figure}
\hbox to\hsize{\epsfxsize=1.0\hsize\hfil\epsfbox{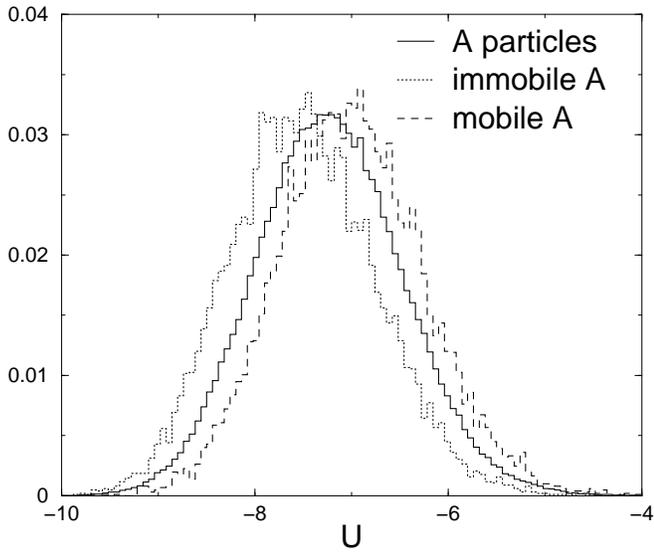}\hfil}
\caption{Distribution of the potential energy of all the A particles, 
of the mobile A particles and of the immobile A particles for $T=0.4510$.}
\label{figene}
\end{figure}

\noindent
\begin{figure}
\hbox to\hsize{\epsfxsize=0.85\hsize\hfil\epsfbox{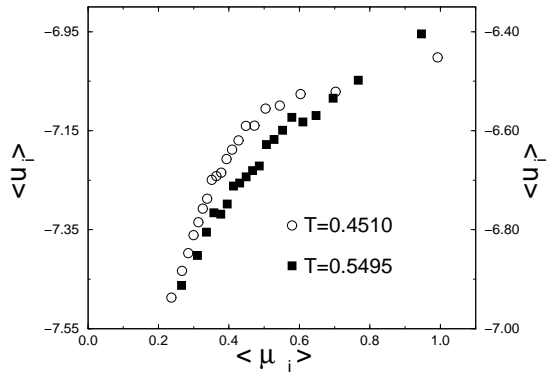}\hfil}
\caption{Potential energy $\langle U_i\rangle$ as a function of the
mobility $\langle \mu_i\rangle$ for the A particles. The A particles
have been divided into 20 subsets according to their mobility at
$t^*$. Each subset is represented by a point in the graph. The
energy scale for $T=0.4510$ is on the left hand side $y$ axis, while
the energy scale for $T=0.550$ is on the right hand side $y$ axis.}
\label{figdistrene}
\end{figure}

\noindent
\begin{figure}
\hbox to\hsize{\epsfxsize=1.0\hsize\hfil\epsfbox{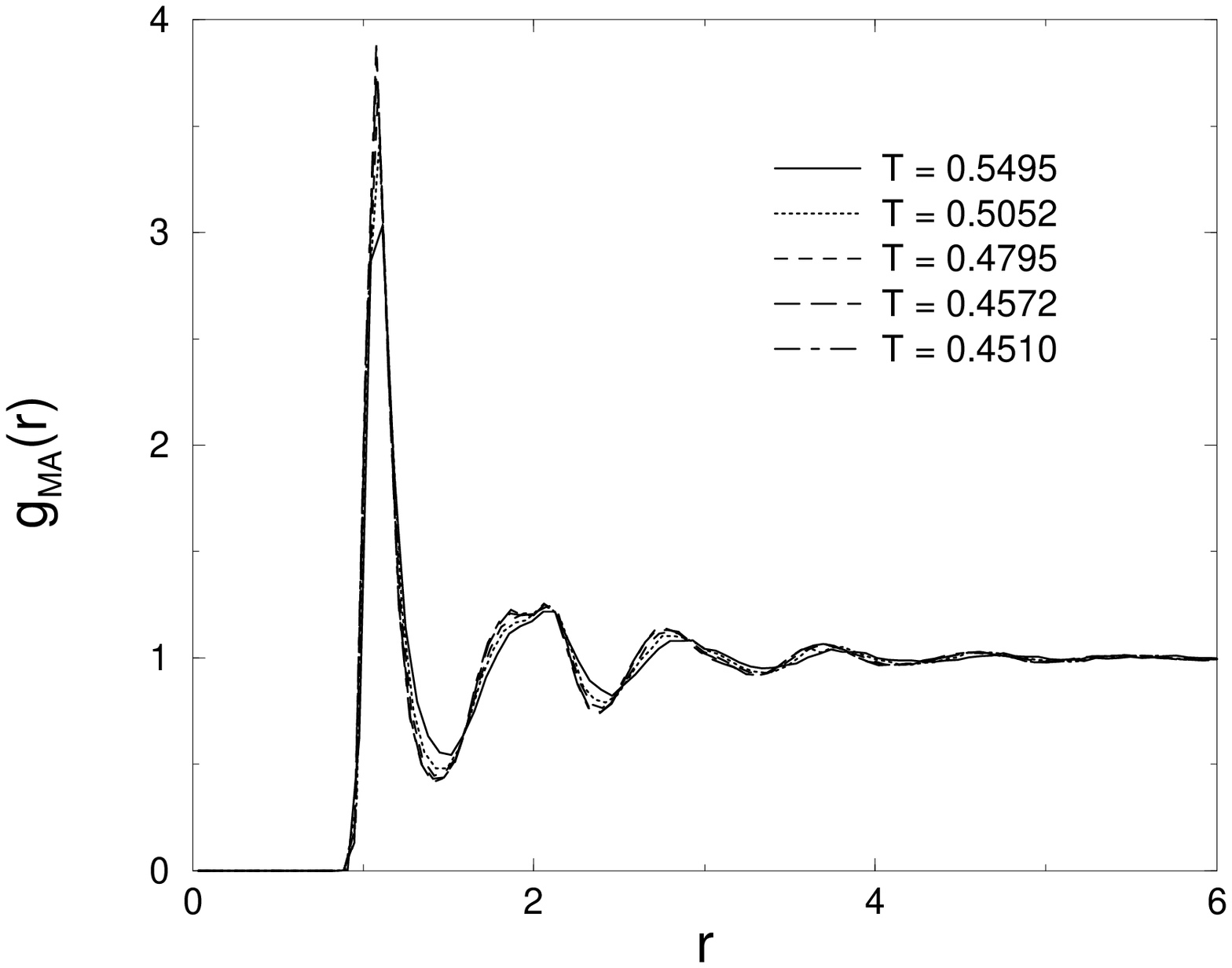}\hfil}
\caption{Pair correlation function $g_{MA}(r)$ between mobile A and generic A particles.}
\label{figgma}
\end{figure}

\noindent
\begin{figure}
\hbox to\hsize{\epsfxsize=1.0\hsize\hfil\epsfbox{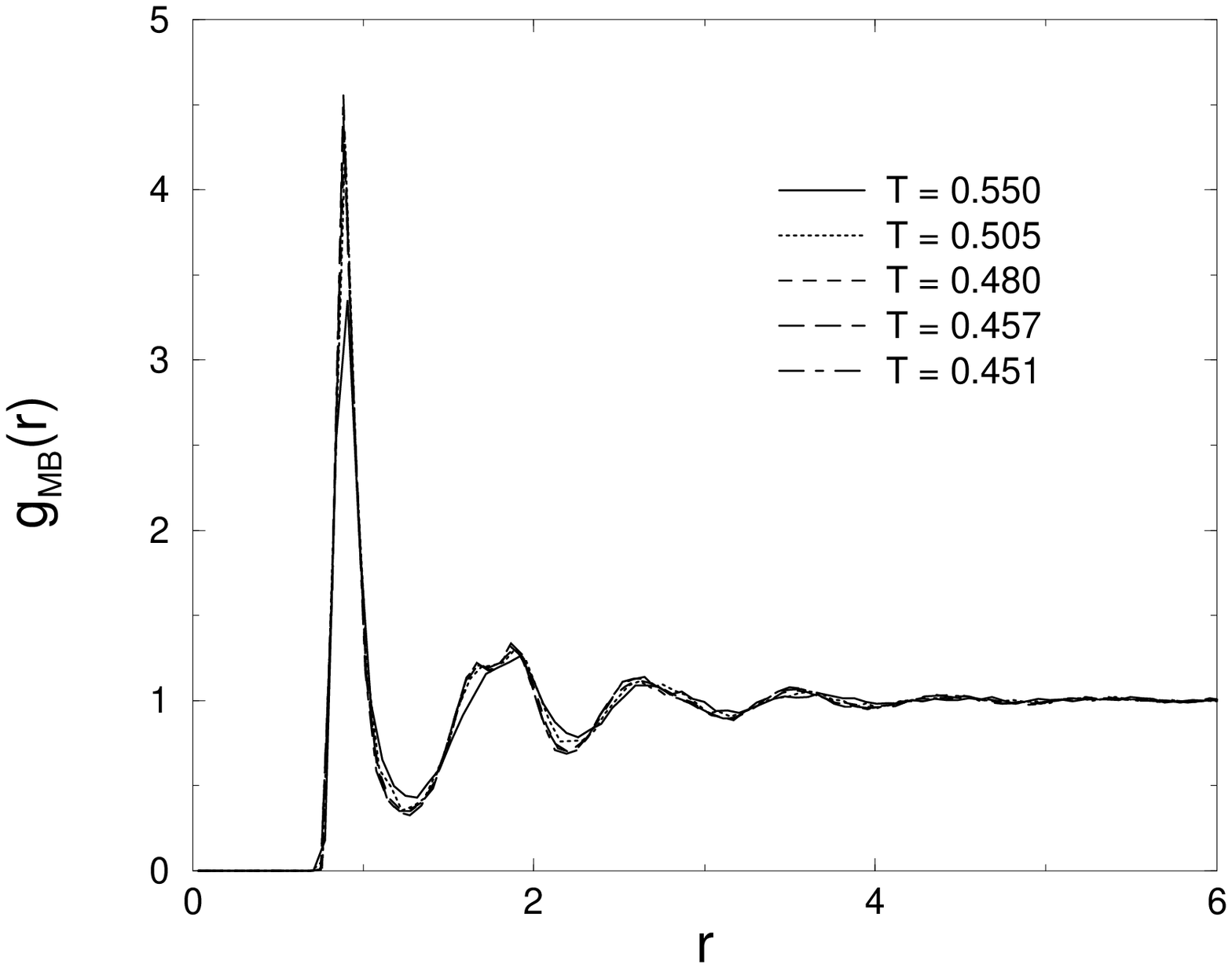}\hfil}
\caption{Pair correlation function $g_{MB}(r)$ between mobile A and generic B particles.}
\label{figgmb}
\end{figure}

\noindent
\begin{figure}
\hbox to\hsize{\epsfxsize=1.0\hsize\hfil\epsfbox{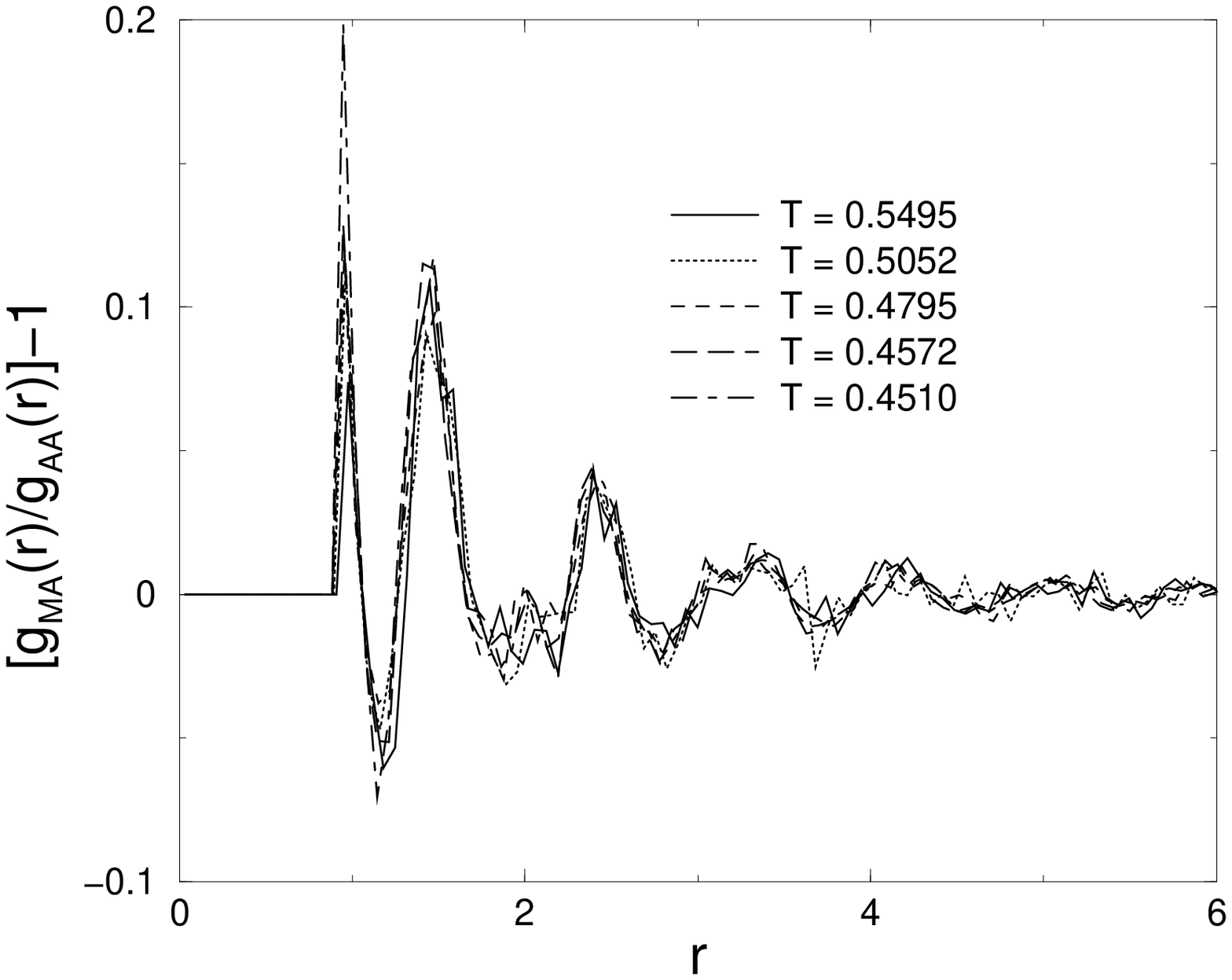}\hfil}
\caption{$\Gamma(r)=[g_{MA}/g_{AA}(r)] -1$ vs. $r$ for different 
temperatures.}
\label{figggmagaa}
\end{figure}

\noindent
\begin{figure}
\hbox to\hsize{\epsfxsize=1.0\hsize\hfil\epsfbox{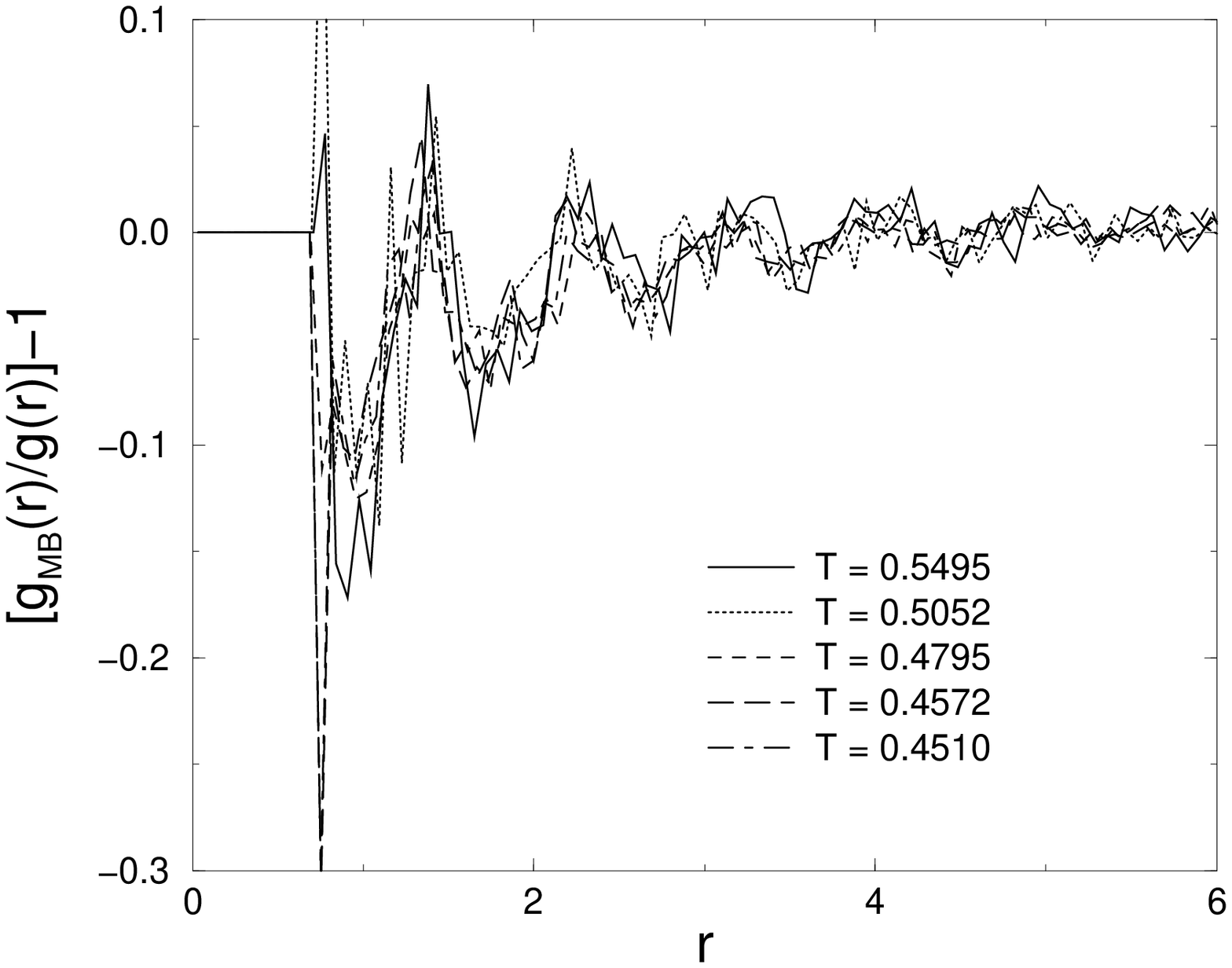}\hfil}
\caption{$\Gamma(r)=[g_{MB}/g_{AB}(r)] -1$ vs. $r$ for different 
temperatures.}
\label{figgmbgab}
\end{figure}

\noindent
\begin{figure}
\hbox to\hsize{\epsfxsize=1.0\hsize\hfil\epsfbox{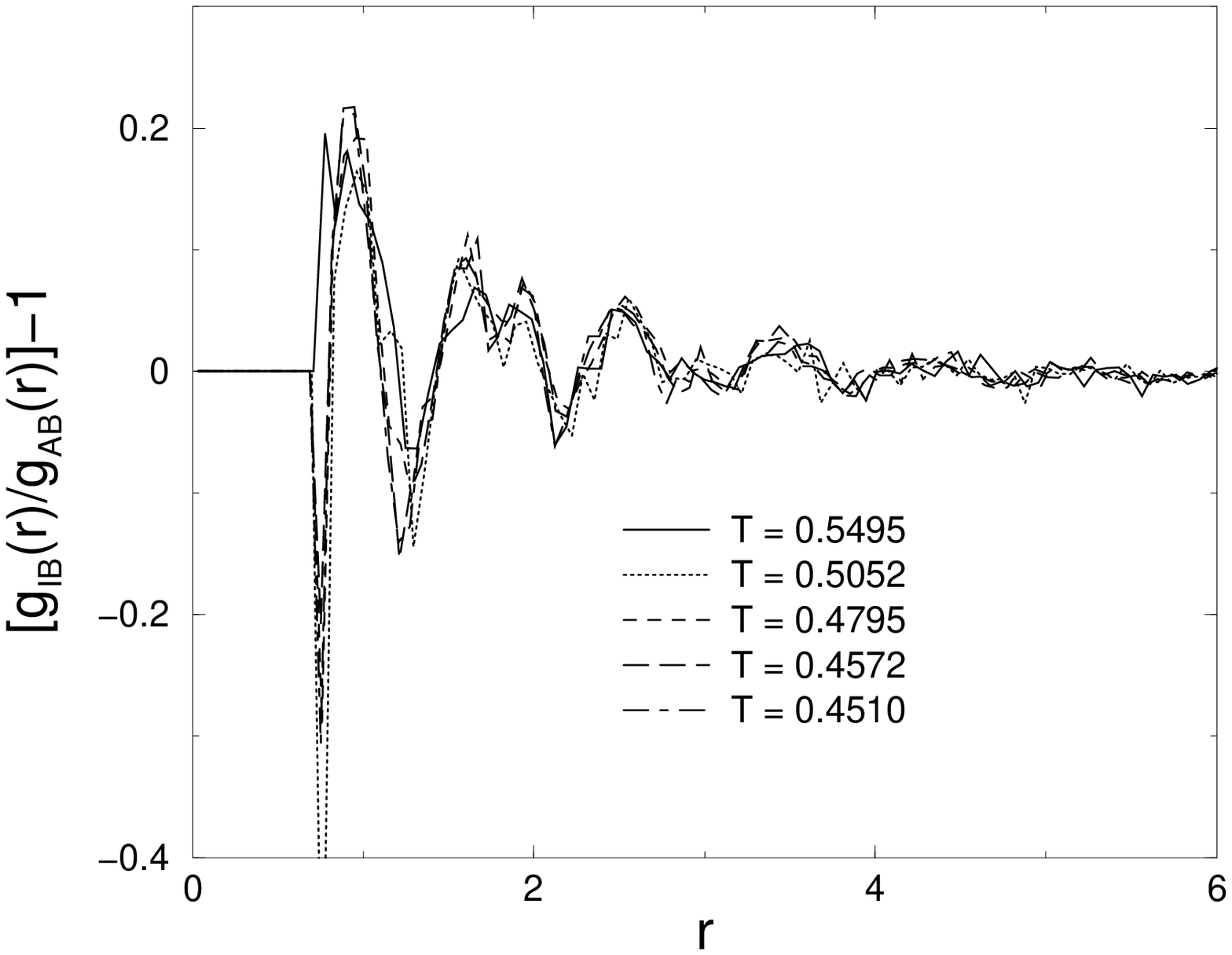}\hfil}
\caption{$\Gamma(r)=[g_{IB}/g_{AB}(r)]-1$ vs. $r$ for different 
temperatures.}
\label{ratio-imm-b}
\end{figure}

\noindent
\begin{figure}
\hbox to\hsize{\epsfxsize=1.0\hsize\hfil\epsfbox{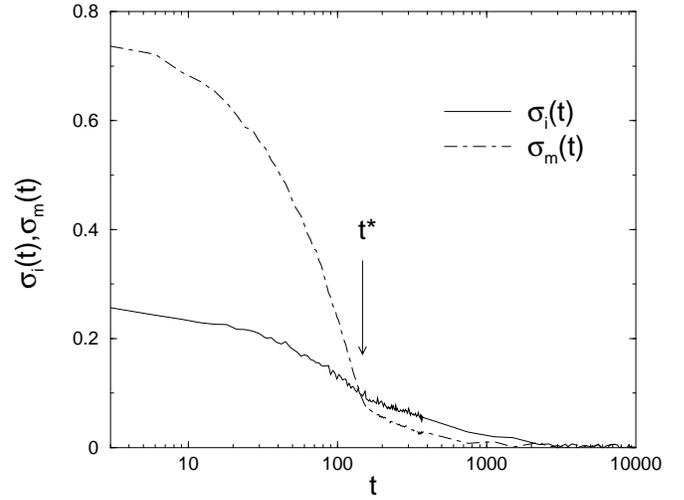}\hfil}
\caption{$\sigma_M(t)$ (dot-dashed line) and $\sigma_I(t)$ (solid
line) for T=0.4510.}
\label{figsigma}
\end{figure}

\noindent
\begin{figure}
\hbox to\hsize{\epsfxsize=1.0\hsize\hfil\epsfbox{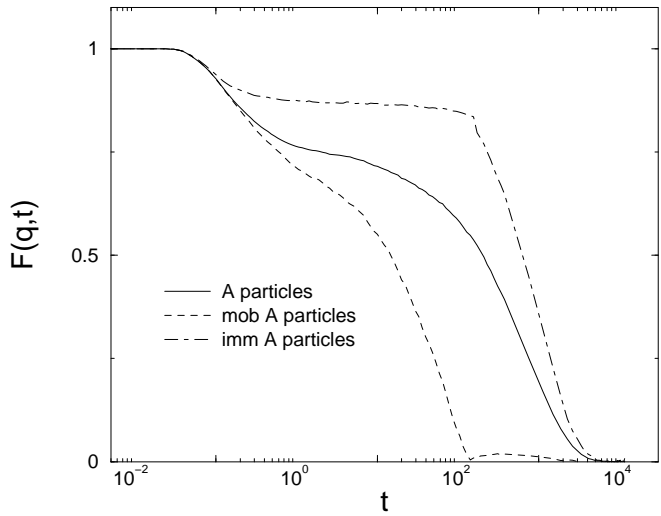}\hfil}
\caption{Self intermediate scattering function for T=0.4510 for all
the A particles (solid line), for the mobile particles (dashed line)
and for the immobile particles (dot-dashed line).}
\label{f_q_t-mob}
\end{figure}

\noindent
\begin{figure}
\hbox to\hsize{\epsfxsize=1.0\hsize\hfil\epsfbox{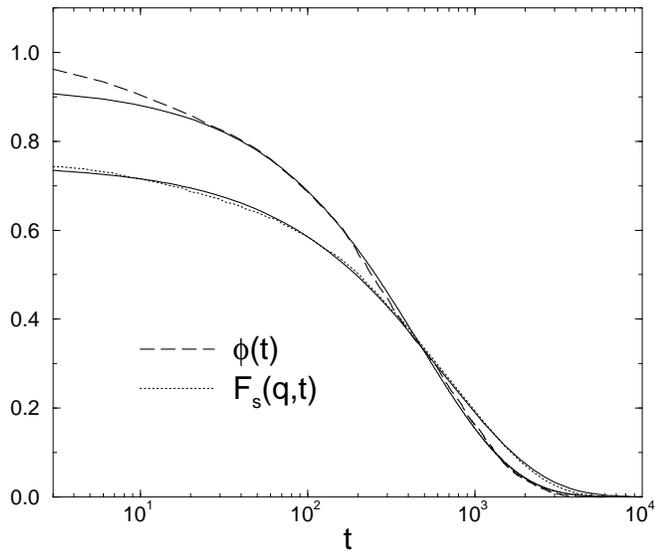}\hfil}
\caption{$\phi(t)$ (dashed line) and $F_s(q,t)$ (dotted line) for
T=0.4510. The solid lines are fits to a stretched exponential.}
\label{gateway}
\end{figure}


\begin{references}
\bibitem{collision}
Here we have defined the ``collision'' time to be
the time when the velocity autocorrelation function first changes
sign.
\bibitem{lengthscale} However, there have been several computational
studies which suggest some type of growing dynamical length scale as
the glass transition is approached.  See, e.g. Ref.~\cite{mountain95,ray}.
\bibitem{immobile} A more promising approach in this direction is to group
particles according to their first passage time; see Ref.~\cite{allegrip}.
\bibitem{return} We have observed that a small fraction of mobile
particles have a tendency to return to their initial positions at
times longer than $t^*$.  These particles may be responsible for this small
memory effect, but further analysis needs to be done.
\end{references}
\end{document}